\documentclass[10pt,journal,twoside]{IEEEtran}

\ifCLASSOPTIONcompsoc
\usepackage[nocompress]{cite}
\else
\usepackage{cite}
\fi

\ifCLASSINFOpdf
\else
\fi

\usepackage{amsmath}
\usepackage{bbm}
\usepackage{enumerate}
\usepackage{booktabs}
\usepackage{multirow}
\usepackage{graphicx}
\usepackage{subfigure}
\usepackage{color}
\usepackage{algorithm}
\usepackage{algorithmic}
\usepackage{hyperref}[colorlinks, linkcolor=black]

\usepackage{graphicx}
\usepackage{latexsym}

\usepackage{times}
\usepackage{soul}
\usepackage{url}

\usepackage{amsthm}
\usepackage{booktabs}

\usepackage[switch]{lineno}
\usepackage{bbding}

\usepackage{graphics,color,multirow,subfigure}
\usepackage{makecell}

\usepackage{amsfonts}

\begin{document}
	%
	\title{Unnoticeable Community Deception via Multi-objective Optimization}
	%
	%
	%
	%
	\author{Junyuan~Fang,~\IEEEmembership{Member,~IEEE}, Huimin Liu, Yueqi Peng, Jiajing~Wu,~\IEEEmembership{Senior Member,~IEEE}, \\Zibin Zheng,~\IEEEmembership{Fellow,~IEEE} and Chi K. Tse,~\IEEEmembership{Fellow,~IEEE}
		\IEEEcompsocitemizethanks{ 
			\IEEEcompsocthanksitem{Junyuan Fang and Chi K. Tse are with the Department of Electrical Engineering, City University of Hong Kong, Hong Kong SAR, China. (Email: junyufang2-c@my.cityu.edu.hk; chitse@cityu.edu.hk)}
			\IEEEcompsocthanksitem{Huimin Liu and Yueqi Peng are with the School of Computer Science and Engineering, Sun Yat-sen University, Guangzhou 510006, China. (Email: liuhm59@mail2.sysu.edu.cn; yiEling27@163.com)}
			\IEEEcompsocthanksitem{Jiajing Wu and Zibin Zheng are with the School of Software Engineering, Sun Yat-sen University, Zhuhai 519082, China. (Email: wujiajing@mail.sysu.edu.cn; zhzibin@mail.sysu.edu.cn)}}
		\thanks{Digital Identifier: IEEE.xxx.xxx.xxxx.xxxxx.}
	}
	%
	%

	\markboth{IEEE Transactions on XXX,~Vol.~XX, No.~XX, XX~2025}%
	{Fang \MakeLowercase{\textit{et al.}}: Unnoticeable Community Deception via Multi-objective Optimization}

	\IEEEtitleabstractindextext{%
		\begin{abstract}
		
		Community detection in graphs is crucial for understanding the organization of nodes into densely connected clusters. While numerous strategies have been developed to identify these clusters, the success of community detection can lead to privacy and information security concerns, as individuals may not want their personal information exposed. To address this, community deception methods have been proposed to reduce the effectiveness of detection algorithms. Nevertheless, several limitations, such as the rationality of evaluation metrics and the unnoticeability of attacks, have been ignored in current deception methods. Therefore, in this work, we first investigate the limitations of the widely used deception metric, i.e., the decrease of modularity, through empirical studies. Then, we propose a new deception metric, and combine this new metric together with the attack budget to model the unnoticeable community deception task as a multi-objective optimization problem. To further improve the deception performance, we propose two variant methods by incorporating the degree-biased and community-biased candidate node selection mechanisms. Extensive experiments on three benchmark datasets demonstrate the superiority of the proposed community deception strategies.

		\end{abstract}
		
		}

	\maketitle

	\IEEEdisplaynontitleabstractindextext

	%
	\IEEEpeerreviewmaketitle

	\section{Introduction}\label{sec:introduction}

	\IEEEPARstart{G}{raphs} or networks, which consist of a series of nodes and links, have been widely employed to model complex relationships in modern social systems.  The typical examples of networks include social networks which illustrate the social relationships between people, financial networks which represent the transaction records between accounts, power networks which indicate the transmission relationship between power nodes, among others \cite{myers2014information,faloutsos1999power,khazane2019deeptrax,pagani2013power}. 
	
	In these years, significant efforts have been put into investigating the abundant information behind graph-based systems, such as community detection for clustering nodes in the graph into different groups, node classification for identifying the labels of nodes, link prediction for predicting possible future connections between nodes, etc \cite{bhagat2011node,lu2011link,khatoon2015survey}. Specifically, in this work, we focus on the task of community detection. Communities have been identified as a widely existing property in many real-world networks. Networks with community structures indicate that nodes in the specific network can be divided into several different groups, where the nodes belonging to the same group share more similar properties and denser links while those nodes belonging to different groups are on the contrary. Currently, numerous studies \cite{newman2004fast,newman2006modularity,blondel2008fast} focusing on community detection tasks have been proposed, helping discover community structures among complex graph data.



	Although current detection algorithms have achieved remarkable success, security concerns have also attracted the attention of researchers in recent years. Taking the social network as an example, from the perspective of individual users, their private and sensitive information may be easily exposed to the public due to these detection tools. Thus, aiming at improving the protection of personal information of entities, or in other words, decreasing the performance of detection algorithms, several community deception strategies \cite{nagaraja2010impact,fionda2017community,waniek2018hiding,chen2019ga} have been proposed. Generally speaking, community deception studies model this problem from the attacker perspective. The goal of attackers is to decrease the performance of detection methods by adding some unnoticeable perturbations to the original network. The studies from the attacker side have shown that current detection algorithms can be easily fooled after the original networks have been slightly modified, such as adding links, removing links, etc.

	However, there are several limitations to current deception strategies. First, in the community detection domain, modularity \cite{newman2006modularity} is one of the most classic metrics to evaluate the community level of networks. Considering this, many studies on community deceptions \cite{chen2019ga,madi2023community,wang2023enhancing} have adopted the decrease of modularity to measure the hiding effect. Nevertheless, based on our empirical study, we interestingly observe that the deception effect does not always follow a simple correlation relationship with the decrease of modularity, indicating that such a measurement may not be a comprehensive metric to characterize the effect of community deception methods.
	
	Moreover, in current deception strategies, a widely adopted assumption is that the attackers are allowed to inject a pre-given budget of perturbations into original networks. However, such a setting limits the flexibility and scalability of attack methods. For instance, in the GAQ method proposed by Chen {\em et al.} \cite{chen2019ga}, attackers need to run the whole algorithm again each time the attack budget setting changes. Therefore, it is necessary to design a more flexible strategy that accommodates different attack budget settings. 
	
	In addition, a successful deception strategy should ensure the similarity of networks before and after the attacks as high as possible. Otherwise, the corresponding perturbations will be easily observed and detected by defenders. However, current attack strategies usually restrict the unnoticeable perturbations to only attacking a limited number of links \cite{fionda2017community,chen2019ga,chen2021community}. We argue that such unnoticeability metrics (i.e., restricting the number of modified links) may also cause a large change to networks. For instance, if we remove a link where one of its connected nodes has only one neighbor, former studies may consider this attack operation as an unnoticeable perturbation since it only removes a single link from the network. However, this operation will make the corresponding connected node become an isolated node, which can be easily detected, indicating that simply limiting the number of attacking links may not be a good measure of unnoticeable perturbations.

	Therefore, to address these issues existing in current deception strategies, this paper proposes an effective and flexible attack strategy to achieve community deception by leveraging evolutionary computation algorithms. By identifying the limitation of current deception metric, specifically the decrease of modularity, we propose a new deception metric based on the adjusted rand index (ARI) \cite{hubert1985comparing}. Moreover, considering the conflict between optimizing the deception performance and attack budget, we model the community deception task as a multi-objective problem (MOP) \cite{deb2016multi,gunantara2018review} to improve the flexibility and scalability of the deception strategy, where the corresponding objectives are the deception effect and attack budget. Specifically, we employ the well-known multi-objective optimization algorithms, NSGA-II \cite{deb2002fast}, as the backbone of our attack method. We then design a degree-preserving mechanism on the crossover and mutation procedures, aiming to achieve unnoticeable perturbations. The proposed mechanism can maintain the degree distribution before and after the attacks as the same, regardless of changes in attack budgets. In particular, a biased mutation operation has been introduced to further improve the deception performance of our method by considering degree-biased and community-biased candidate node selections. Extensive experimental results demonstrate the superiority of the proposed deception framework across several benchmark datasets. The main contributions of this work are summarized as follows.

	\begin{enumerate}
		\item We empirically identify the limitations of the current community deception metric, i.e., the decrease of modularity, and further adopt a new comprehensive deception metric based on adjusted rand index (ARI) instead.
		\item We model the problem in this study as a multi-objective optimization task by considering the deception performance and attack budget simultaneously to improve the flexibility and scalability of community deception strategies.
		\item We develop an unnoticeable deception strategy based on multi-objective evolutionary computation methods, which maintains the degree information of each node before and after the attacks. Moreover, two variant strategies that consider the biased mutation process have been developed to further enhance the deception performance. 
		\item We conduct comprehensive experiments to evaluate the superiority of the proposed community deception strategies across several benchmark datasets.
	\end{enumerate}

	The rest of our paper is organized as follows. Section \ref{sec:related} introduces the related work on community detection and community deception. Two specific metrics and their relationship are presented in Section \ref{sec:metrics}. Section \ref{sec:method} presents the detailed methods and Section \ref{sec:exp} demonstrates the experimental results. Finally, we provide a brief conclusion in Section \ref{sec:con}.

	\section{Related Work}\label{sec:related}
	In this section, we review related work in the areas of community detection and community deception.
	
    \subsection{Community Detection}
	Community detection stands as a cornerstone in the analysis of complex networks, enabling the identification of dense structures that often correspond to functional or social groupings within a broad range of networks. A large number of approaches for community detection have been proposed since the problem was first introduced, such as topology-based methods \cite{blondel2008fast,girvan2002community,ng2001spectral}, statistical model-based methods \cite{holland1983stochastic,airoldi2008mixed}, machine learning-based methods \cite{yang2016modularity,cavallari2017learning,zhang2020commdgi}. As our work focuses on the topology-based community detection, we will mainly review the representative works in this category.

	Girvan and Newman \cite{girvan2002community} first proposed a simple yet effective method to detect communities by progressively removing links with the highest betweenness from the original graph, assuming that such links play crucial roles in the networks. After that, Raghavan {\em et al.} \cite{raghavan2007near} proposed a near linear time detection algorithm (LPA) by leveraging the label propagation. Modularity is a widely adopted metric to measure the community level of a specific network. As it is almost impractical to directly maximize the modularity to discover communities due to high computational costs, Blondel {\em et al.} \cite{blondel2008fast} proposed the powerful and scalable Louvain algorithm (LOU)  by dividing the modularity optimization problem into local and global levels. Furthermore, Newman proposed a spectral optimization algorithm (SOA) \cite{newman2006finding} by using the spectral properties, such as eigenvalues and eigenvectors of the Laplacian matrix of networks. Lots of efforts have been made from different perspectives to discover the potential community structures in networks, such as graph cuts \cite{wang2015community,shin2022graph}, game theories \cite{zhou2013game,jiang2015dynamic,chen2019community}, random walks \cite{avron2015community,hollocou2016improving}, local structures \cite{maity2014extended,baudin2022clique}, etc. With the continuous improvement of detection performance, concerns about the leakage of personal privacy have grown, which facilitates the development of community hiding studies.

	\subsection{Community Deception}
	Community deception, also known as community hiding, is an inverse problem of community detection. It deliberately introduces ambiguity or misinformation into a graph-based representation of data to avoid community detection. The idea of community deception was first introduced by Nagaraja \cite{nagaraja2010impact} in 2010, which countered surveillance in a real-world social network. Later, Waniek {\em et al.} \cite{waniek2018hiding} proposed the DICE mechanism (i.e., disconnect internally and connect externally), which disconnects the existing node pairs between the same cluster and connects the node pairs between different clusters, to influence the downstream detection task. As the discussion on this issue deepens, this topic has begun to attract more attention. Typically, the works in this field can be divided into two main types, namely focusing on misleading a target community and decreasing the overall detection of entire graphs. 
	
	From the perspective of misleading a target community, Fionda {\em et al.} \cite{fionda2017community} first proposed a safeness-based deception strategy considering the reachability preservation, community spread, and community hiding of the target community at the same time. Chen {\em et al.} \cite{chen2021community} further proposed an improved gain function to enhance the effect and efficiency of the deception algorithm by only focusing on inter-community link additions and intra-community link deletions. Based on the proposed gain function, Yu {\em et al.} \cite{yu2021safeness} developed a new deception strategy by utilizing the multi-simulated annealing algorithm. Moreover, Chen {\em et al.} \cite{chen2020multiscale} developed a new fitness function that considers the change in degree information before and after attacks as the unnoticeable constraint by utilizing evolutionary computation methods. Recently, aiming to achieve a black-box attack where the specific detection algorithm is unknown, Yang {\em et al.} \cite{yang2023lsha} proposed a local structure-based deception approach that focuses on some local nodes with dense connections instead of the whole community structures. Chang {\em et al.} \cite{chang2024community} investigated the complete escape problem of the target community by utilizing the escape scores, dispersion scores, and hiding scores as the evaluation metrics using genetic algorithm.
	
	On the other hand, from the perspective of decreasing the global detection performance, Chen {\em et al.} \cite{chen2019ga} first proposed the GAQ attack method by leveraging the metric of the decrease of modularity through genetic algorithm. Besides misleading the target community, the study \cite{chen2020multiscale} also proposed another fitness function that focuses on the global deception performance. Moreover, by observing the Matthew effect that exists in the allocation of perturbation resources, Liu {\em et al.} \cite{liu2021prohico} proposed a probabilistic framework by incorporating the random allocation and likelihood minimization to mitigate the imbalanced budget allocation issue. Unlike the study \cite{chen2019ga} that adopts modularity as the evaluation metric, Liu {\em et al.} \cite{liu2022hiding} selected the normalized mutual information (NMI) as the corresponding metric and further employed genetic algorithm to obtain the optimal perturbations. Liu {\em et al.} \cite{liu2022community} proposed the GCH deception method by determining the optimal perturbations based on the prediction confidence provided by a trained graph autoencoder \cite{kipf2016variational} . To improve the scalability of evolutionary computation-based methods in large-scale networks, Zhao {\em et al.} \cite{zhao2023obfuscating} first developed a divide-and-conquer strategy and then proposed a co-evolutionary optimization algorithm. Furthermore, Zhao {\em et al.} \cite{zhao2023self} proposed another attack strategy by constraining the modification distances between the potential perturbations to reduce the search space and improve the attack performance simultaneously. In addition, to improve the attack performance in large-scale networks, Liu {\em et al.} \cite{wang2023enhancing} first sampled several sub-graphs from the original graph, and then employed a graph autoencoder to generate the possible perturbations for each sub-graph, respectively, where the perturbations from each sub-graph will finally be determined based on genetic algorithm. Moreover, Liu {\em et al.} \cite{liu2024unified} determined the optimal perturbations based on the guidance of nonnegative matrix factorization on the adjacency matrix.
	
	Besides the above two perspectives, there are also several works that focus on other aspects, such as the community hiding in the attributed graphs \cite{li2020adversarial}, the hiding of nodes in overlapping communities \cite{liu2022protect,yang2022overlapping,ji2024micro}, etc. Although a great deal of studies have been proposed to investigate the community deception problem, several limitations exist. First, from the perspective of misleading the entire network, one of the widely used evaluation metrics is the decrease of modularity. However, we empirically observe that it may be insufficient to characterize the deception performance. Second, most current deception strategies lack flexibility and scalability because they need to pre-define a specific attack budget. Third, few of the aforementioned works consider the issue of unnoticeable perturbations, as most of them typically pre-define a fixed budget as the unnoticeable constraint. However, it is crucial to maintain the overall unnoticeability of the deception strategy, otherwise such attacks may be easily identified by detection systems. Thus, in this work, we are dedicated to addressing the above limitations in current hiding strategies by proposing a degree-preserving multi-objective community deception strategy.

	\begin{figure}[t]
		\centering

		\includegraphics[width=0.8\linewidth]{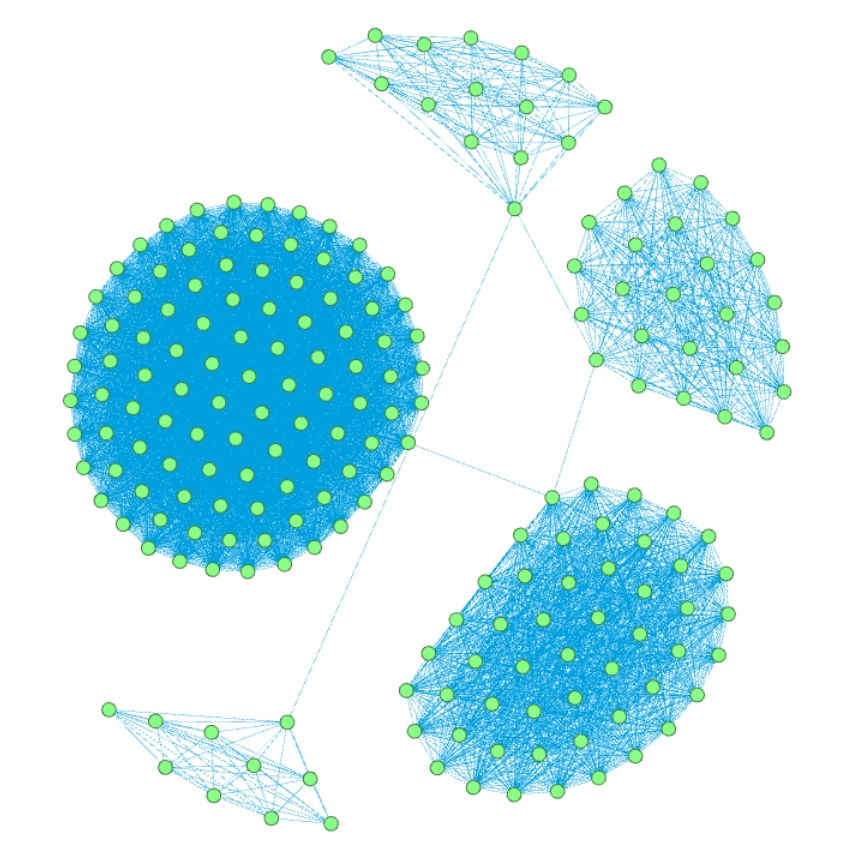} 

		\caption{Illustration of the generated graph with a strong community structure containing 200 nodes, and the community distribution is \{100, 50, 25, 15, 10\}. Here the green circles and blue lines represent the nodes and links in the graph, respectively.}
		\label{fig:illustration}
	\end{figure}

	\begin{table*}[h]
		\centering
		\renewcommand{\arraystretch}{1.2}
		\caption{Comparisons of the influence of different community adjustments, such as merging or splitting the large or small communities, on the corresponding community metrics under the generated networks with 200 and 400 nodes. Here, the communities represent the specific community distributions. $Q$ and ARI denote the corresponding modularity and adjusted rand index metrics.}
		\resizebox{0.8\linewidth}{!}{
		\begin{tabular}{c||c|c|c||c|c|c}
			\hline\hline
			\multirow{2}{*}{Operation}&
			\multicolumn{3}{c||}{$n$=200} & \multicolumn{3}{c}{$n$=400} \\
			
			\cline{2-7} & Communities & $Q$ & ARI &   Communities & $Q$ & ARI \\
			\hline\hline
			Original & \{100,50,25,15,10\} & 0.4051 & - & \{200,100,50,30,20\} & 0.4077   & -\\
			\hline
			Merge large & \{150,25,15,10\} & 0.0753 & 0.5123 & \{300,50,30,20\} & 0.0769   & 0.5255\\
			\hline
			Split small & \{100,50,25,15,5,5\} & 0.4009 & 0.9831 & \{200,100,50,30,10,10\} &  0.4033 &  0.9859\\
			\hline
			Merge large \& split small& \{150,25,15,5,5\} & 0.0715 & 0.5123 & \{300,50,30,10,10\} &  0.0730 & 0.5255\\
			\hline
			Merge small & \{100,50,25,25\} & 0.4295 & 0.9831 & \{200,100,50,50\} &  0.4317  & 0.9832\\
			\hline
			Split large & \{50,50,50,25,15,10\}  & 0.7284 & 0.6876 & \{100,100,100,50,30,20\} &  0.7300 & 0.6897 \\
			\hline
			Merge small \& split large& \{50,50,50,25,25\} & 0.7430 & 0.6710 & \{100,100,100,50,50\} & 0.7442 & 0.6731  \\
			\hline
			\bottomrule
		\end{tabular}}
		\label{tab:community_dis}
	\end{table*}

	\section{Preliminaries}\label{sec:metrics}
	In this section, we first introduce the general representations of graphs. Then, we discuss the limitations of simply employing the decrease of modularity to measure the deception performance. In addition, we introduce a more appropriate deception metric that we adopt in this work, along with the metric for attack budgets. Finally, we study the correlations between two proposed metrics to verify the rationality of the modeling of multi-objective optimizations.

	\subsection{Graph Representations}
	Generally speaking, a graph can be represented as a collection of nodes and their corresponding links, i.e., $G= \{V, E\}$, where $V = \{v_1, v_2, \cdots, v_n\}$ and $E = \{e_1, e_2, \cdots, e_m\}$ denote the set of nodes and links in the graph $G$, respectively. Moreover, $A$ is the adjacency matrix representing the connection relationship between nodes, $A_{i,j} = 1$ if there is a link between nodes $v_i$ and $v_j$, and 0 otherwise.

	\subsection{Limitations of the Decrease of Modularity}

	Modularity is a widely used metric to evaluate the performance of community detection methods, where a higher modularity usually can be considered that there is a better community structure in the corresponding network \cite{newman2006modularity}. Thus, it is a natural idea to employ the decrease of modularity to measure the performance of community deception in lots of previous studies. However, in our empirical study, we interestingly observe that a good deception performance does not always indicate the decrease of modularity. On the contrary, sometimes it may lead to an increase of it. 
	
	To verify the comprehensiveness of the decrease of modularity metric, we design a graph generator for synthetic networks with strong community structure. Specifically, for a given community distribution within a network, we first generate a fully connected subgraph for each community. Then, each nearby community is connected in a chain-like configuration via a single link. For example, a graph with 200 nodes and five communities, i.e., \{100, 50, 25, 15, 10\}, is visualized in Fig. \ref{fig:illustration}. 
	
	By adopting the proposed graph generator, we then explore the relationship between the change of community distribution and modularity, as shown in Table \ref{tab:community_dis}. We take the graph with 200 nodes (i.e., $n$=200), whose original community distribution is \{100, 50, 25, 15, 10\}, as an example. Six different operations have been conducted to mimic the attack operation at an extreme level, including the merge of the two largest or smallest communities, the split of the largest or smallest community, and their combinations. Each of them can be considered as a strong attack operation. For instance, the merge of communities with 100 and 50 nodes is considered to be the merge of the two largest communities. Under this operation, the nodes in these two communities will be reconstructed as a new fully connected subgraph. Although the nodes that originally belonged to the same community will still be predicted to belong to the same community, they will also be mispredicted to the same community as nodes from different communities originally. Conversely, the split of the community with 10 nodes will be considered as the split of the smallest community. Under this operation, the nodes in this community will be divided into two new communities where each has 5 nodes and the nodes within each new community will be fully connected. Although the 5 nodes in each new community will still be predicted to be the same community as each other, they will be mispredicted as different from the other 5 nodes that belonged to the same community originally. Therefore, the six operations above can all be considered as strong attack operations.

	However, by observing the results presented in Table \ref{tab:community_dis}, we find that not all six attack operations correspond to the decrease of modularity. Merging the large communities, splitting the small community, and their combinations tend to lead to the decrease of modularity, while merging the small communities, splitting the large community, and their combination tend to lead the increase of modularity. Through reviewing the specific operations, we find that high modularity requires not only a strong community structure but also a balanced community scale, as splitting a large community or merging small communities both help balance the community scale. We also obtain a similar conclusion on a larger network with $n$=400. Based on the above results, it is obvious that the decrease of modularity is not a comprehensive metric to measure the deception performance.

	\subsection{Measure of Attack Performance}
	
	To evaluate the performance of the attack strategy, instead of the decrease of modularity adopted by most previous studies, we employ the concept of adjusted rand index (ARI) \cite{hubert1985comparing} in this work. ARI is another widely used metric that characterizes the performance of community detection algorithms. Specifically, ARI considers all pairs of nodes that are assigned to the same or different clusters in the predicted and true labels. One of the advantages of ARI is that it is not sensitive to the size of clusters, meaning it is not biased towards networks with clusters having similar sizes. 
	
	Specifically, we first employ a specific detection method on the clean graph (i.e., before the attacks) and consider the obtained clustering labels as the ground truth labels. Our goal is to design the attack strategy to mislead nodes to be predicted to their corresponding ground truth labels. In particular, to further transform our task into a maximization problem, we define the measure of attack performance as the decrease of ARI, namely DARI, which is given as follows.
	
 	\begin{equation}\label{eq:dari}
	 	{\rm DARI} = 1 - {\rm ARI},
	\end{equation} 
	where 1 indicates the original ARI value of the clean graph before the attacks. A larger value of DARI represents a better deception performance.
	
	\subsection{Measure of Attack Budgets}
	For the metric of attack budgets, we simply denote it by utilizing the links modified during attacks. The details are as follows.
	
	\begin{equation}\label{eq:at}
		{\rm AT} = \frac{|A^{\prime} - A|}{2},
	\end{equation} 
	where $A^{\prime}$ and $A$ denote the adjacency matrices of the graph after and before attacks, respectively. An additional denominator of 2 is included as both $A$ and $A^{\prime}$ are symmetric matrices. Moreover, to transform our task into a maximization problem, we further define the decrease of AT, denoted as DAT, as follows.
	\begin{equation}\label{eq:dat}
		{\rm DAT} = 1 - \frac{|A^{\prime} - A|}{2T},
	\end{equation} 
	where T denotes the largest perturbation budget, and we set it at 20\% of the number of original links. Generally, a larger value of DAT represents a lower perturbation budget.

	\begin{figure*}[t]
		\subfigure[\textbf{Karate}]{
			\begin{minipage}[]{0.33\linewidth}
				\includegraphics[scale=0.33]{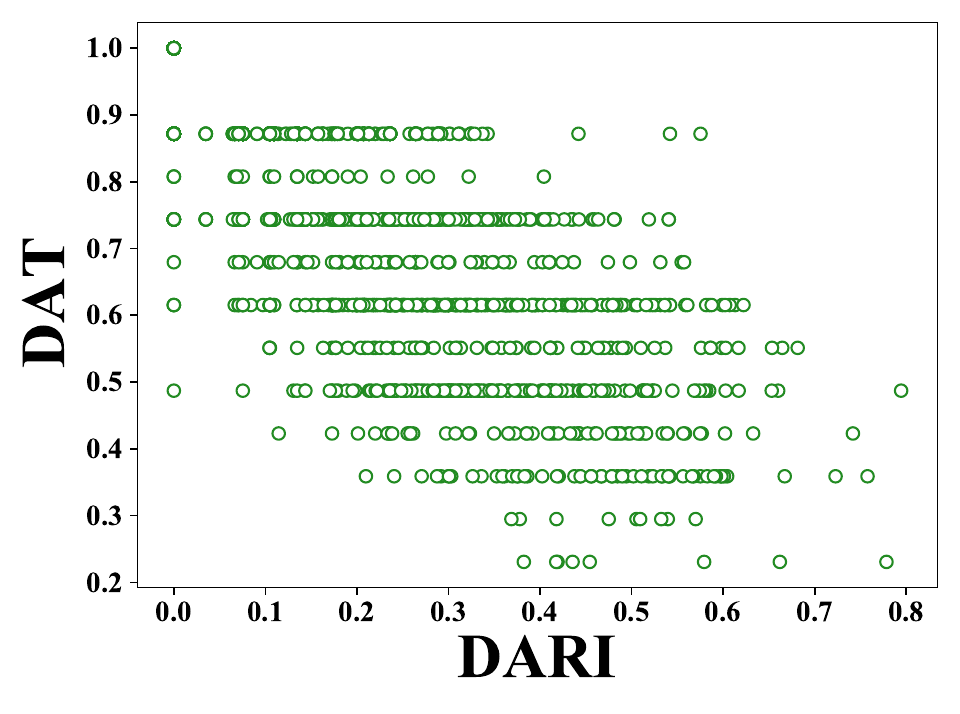}
			\end{minipage}%
		}%
		\subfigure[\textbf{Dolphins}]{
			\begin{minipage}[]{0.33\linewidth}
				\includegraphics[scale=0.33]{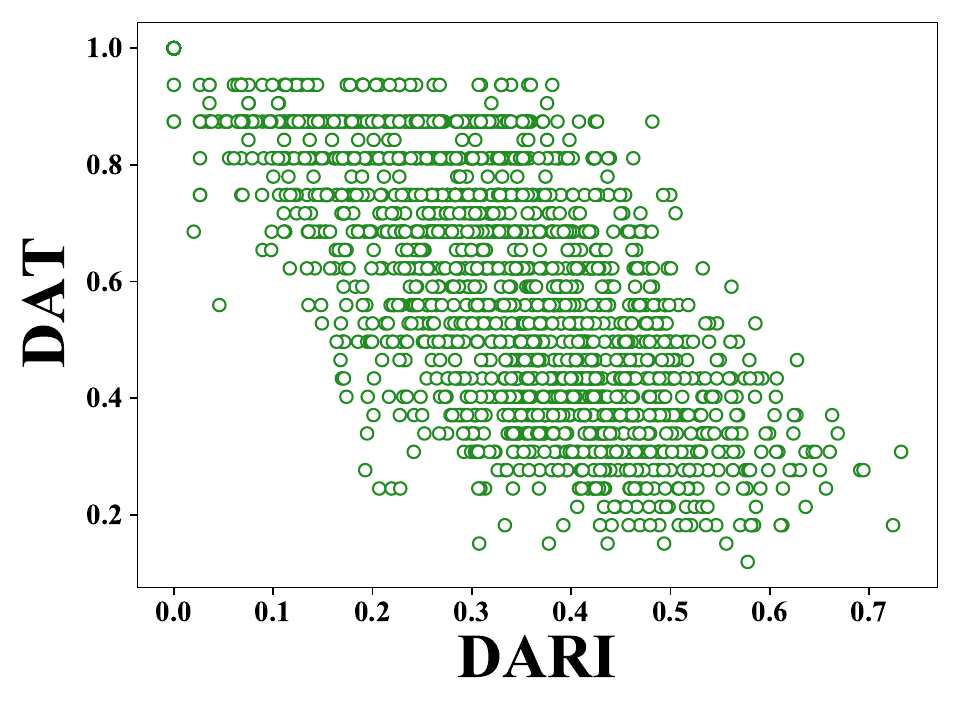}
			\end{minipage}%
		}%
		\subfigure[\textbf{Netscience}]{
			\begin{minipage}[]{0.33\linewidth}
				\includegraphics[scale=0.33]{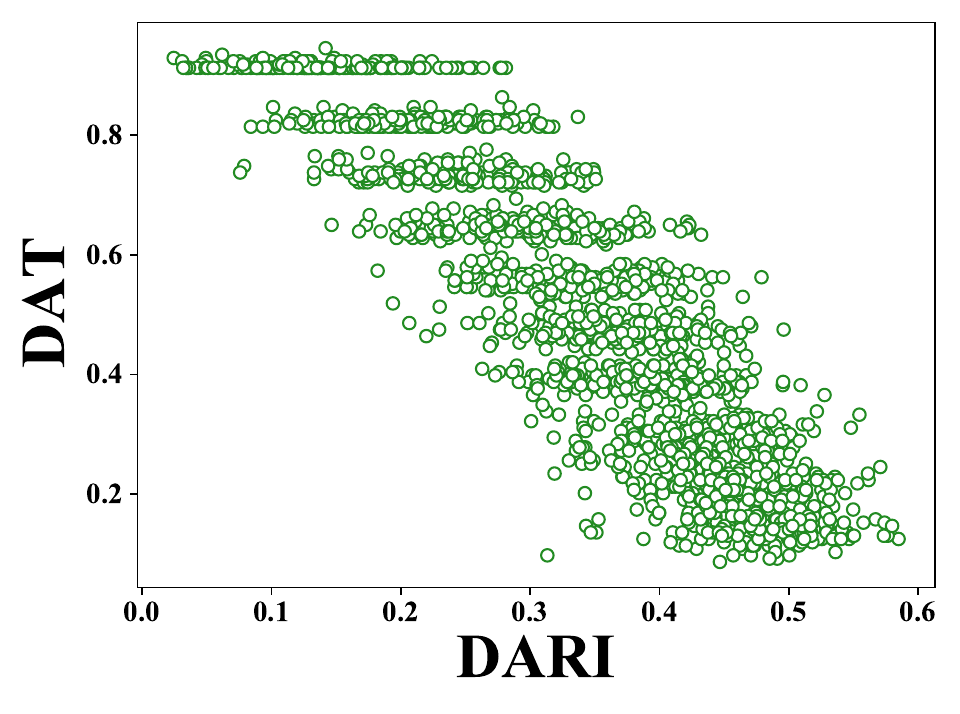}
			\end{minipage}%
		}%

		\centering
		\caption{Visualization of the correlation between the DAT and DARI metrics in three datasets using the LOU detection method based on 2,000 independent simulation runs. Each circle demonstrates a specific attack operation on the corresponding network.}
		\label{fig:cor}
	\end{figure*}

	\subsection{Correlation Study}
	Before the proposing of our attack strategy, we first explore the correlation of the proposed two evaluation metrics, i.e., DARI and DAT. Specifically, we sequentially increase DAT (i.e., decrease AT) and then record the corresponding DARI. We randomly generate 2,000 independent perturbations for visualization. The correlation of the two metrics for the three benchmark datasets under the aforementioned operations is given in Fig. \ref{fig:cor}. Particularly, the statistics for the three datasets are given in Table \ref{table:dataset}.
	
	From Fig. \ref{fig:cor}, we can observe that the increase in DAT tends to decrease DARI. In other words, a better deception performance usually requires a larger attack budget. The above result is reasonable since attackers can allocate more resources to influence more nodes to be misclassified into the wrong communities when DAT decreases. Therefore, to achieve a better deception performance while using a lower attack budget, it is both reasonable and necessary to model the optimization of the above two conflict metrics as a multi-objective optimization problem.

	\section{Proposed Method}\label{sec:method}
	In this section, based on the discussion in the previous section, we introduce the proposed attack strategies focusing on maximizing DARI and DAT simultaneously. Specifically, we first present the details of the proposed unnoticeable method designed based on the backbone of the NSGA-II \cite{deb2002fast} method step by step. Then, we introduce two improved strategies by inducing the biased mutation mechanism. Finally, we give an overall framework for community deception via multi-objective optimization.
	
	\begin{figure}[t]
		\centering
		\includegraphics[width=0.75\linewidth]{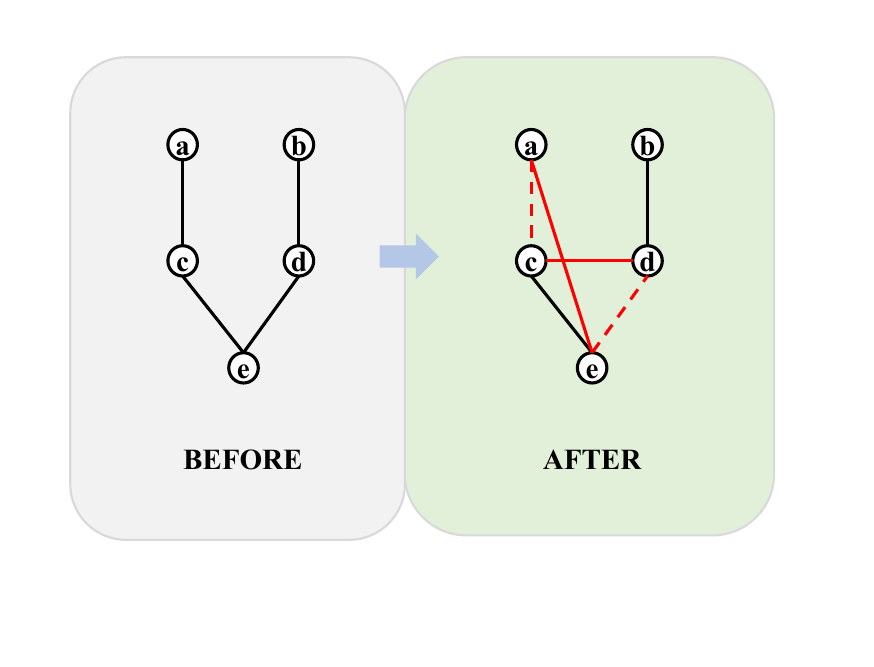}
		\vspace{-10pt}
		\caption{Illustration of the degree-preserving rewiring operation (and mutation process), where solid lines and dashed lines represent the added links and removed links, respectively.}
		\label{fig:mutation}
	\end{figure}

	\begin{figure}[t]
		\centering
		
		\includegraphics[width=0.8\linewidth]{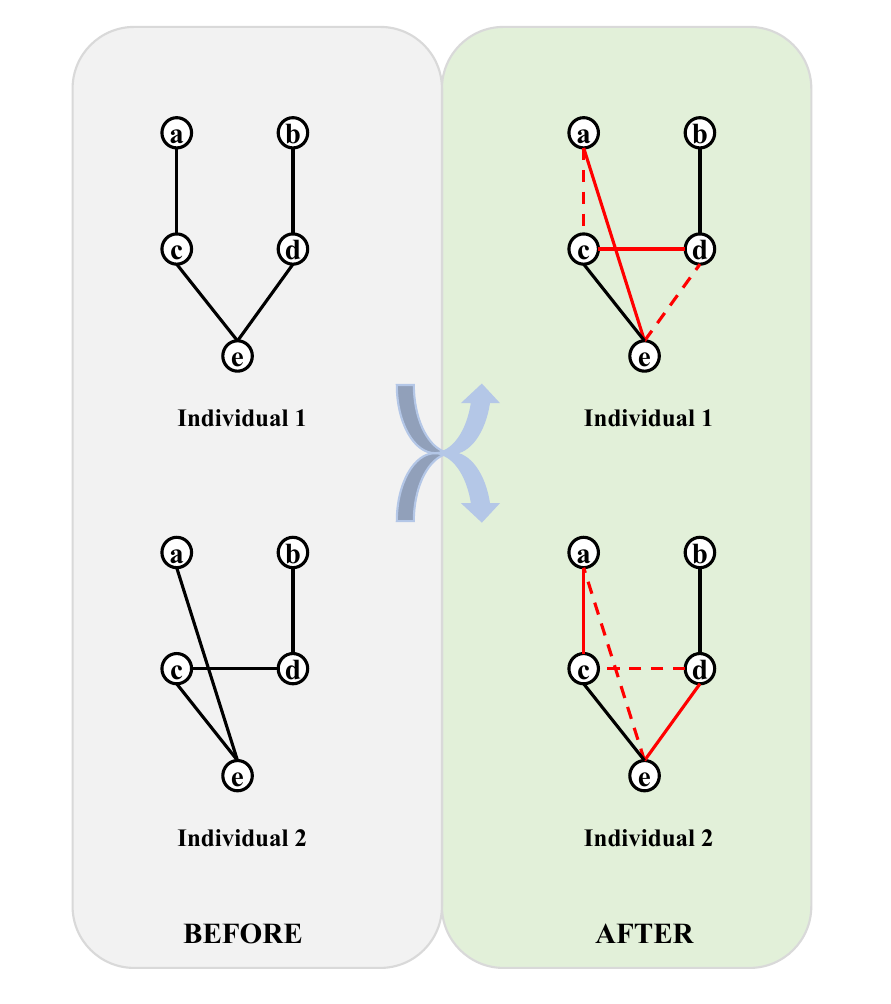} 
		
		\caption{Illustration of the crossover process, where solid lines and dashed lines represent the added links and removed links, respectively.}
		\label{fig:crossover}
	\end{figure}

	\begin{figure*}[]
		\centering
		
		\includegraphics[width=0.8\linewidth]{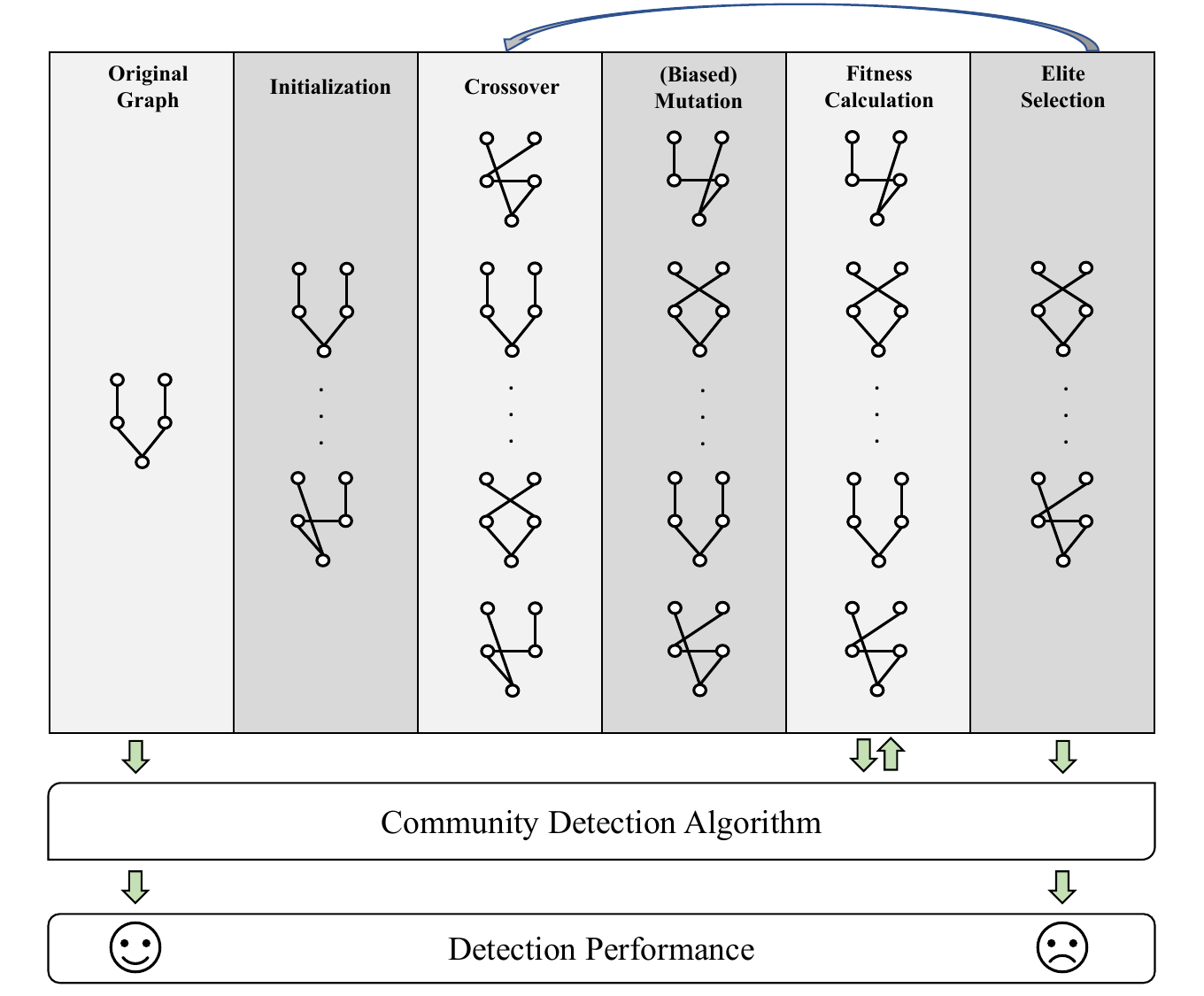} 
		
		\caption{A systematic framework of UCD-based methods, where the biased mutation represents the corresponding variant methods of the original UCD method by inducing the degree-biased and community-biased candidate node selection mechanisms.}
		\label{fig:framework}
	\end{figure*}

	\subsection{Degree-preserving Multi-objective Optimization}
	Based on the above discussion, we can conclude that the two investigated metrics, i.e., DARI and DAT, are in conflict with each other. Thus, we further model the optimization of these two metrics as a multi-objective problem. Given a specific graph $G$, our goal is to maximize both DARI and DAT by injecting unnoticeable perturbations. Specifically, we adopt the backbone of the NSGA-II method to design our multi-objective \underline{u}nnoticeable \underline{c}ommunity \underline{d}eception method, i.e., UCD, as NSGA-II demonstrates both low computational cost and fast convergence speed.
	
	The detailed process of UCD for addressing our problem is given as follows.
	
	{\em \textbf{Step 1: Initialization.}} Copy the original $G$ and then randomly perturb it to generate a new ``individual", denoted as $G_i$. Specifically, we adopt an unnoticeable rewiring operation to inject perturbations into the graph. A simple example for illustrating the rewiring operation is given in Fig. \ref{fig:mutation}. By selecting node $c$ as the target node, we first disconnect one of its original links (i.e., link $a-c$) and connect a new link (i.e., link $c-d$) that does not exist in the original graph. After that, to achieve the rewiring operation, we disconnect one of the original links of node $d$ where the end node is not connected to node $a$ (i.e., link $d-e$). Finally, we connect a new link (i.e., link $a-e$) to complete the rewiring operation. It is worth noting that the degree of each node after the rewiring operation will remain the same as that before the attacks. Thus, each individual can be considered as a possible unnoticeable attack by injecting malicious perturbations via the rewiring operation. This individual generation process will be repeated $\Omega$ times to initialize different individuals by employing different times of rewiring operation executions. Finally, all individuals constitute the population $P$.

	{\em \textbf{Step 2: Crossover.}} Among $\Omega$ different individuals in $P$, we randomly select two individuals, i.e., $G_i$ and $G_j$, and randomly exchange a part of their information to achieve the individual crossover operation under a crossover probability $p_c$. Taking Fig. \ref{fig:crossover} as an example, we assume the two individuals in the crossover operation are individuals 1 and 2. We first randomly determine a target node that will share a certain part of information with the corresponding node in the other individual, for instance, node $a$ in our example. After that, we can obtain the neighbor sets of node $a$ (i.e., \{$c$\} and \{$e$\}) in the two individuals, respectively. Then, node $a$ in individual 1 will disconnect one of its original neighbors (i.e., link $a-c$) and connect one of the neighbors of node $a$ in individual 2 (i.e., link $a-e$). In addition, to maintain the degree distribution of individual 1, we further disconnect one of the neighbors of node $e$ (i.e., link $d-e$) and connect the original neighbor of node $c$ to the neighbor of node $e$ (i.e., link $c-d$). After that, the degree distributions of individual 1 before and after the crossover operation remain the same. At the same time, we adopt a similar operation on individual 2 as well. Through the above crossover operation, the two newly generated individuals will contain certain information from each other while constraining the degree distribution of the individuals to remain the same before and after the attacks.
	
	{\em \textbf{Step 3: Mutation.}}	Different from the crossover operation, which focuses on exchanging information between different individuals, the mutation operation aims to induce some randomness in the population evolution. The specific mutation process is similar to \textbf{\em Step 1}. Under a pre-defined mutation probability $p_m$, a target node will first be determined. Then, a degree-preserving rewiring operation will be employed, as illustrated in Fig. \ref{fig:mutation}.
		
	{\em \textbf{Step 4: Fitness Calculation.}} After the crossover and mutation steps, for each individual in the population $P$, we input the specific denoted graph into the targeted community detection methods and record the corresponding metrics, namely, the deception performance DARI and the decrease of attack budget DAT. 
	
	{\em \textbf{Step 5: Non-dominated Sorting and Elite Selection.}} Because of the crossover operation, the scale of the original $P$ increases from $\Omega$ to $2\Omega$. Therefore, it is necessary to reduce the scale of $P$ back to $\Omega$ and only reserve those individuals with better fitness scores. By giving a series of individuals with different DARI and DAT, our goal is to determine which of them should be reserved. Specifically, a non-dominated sorting is adopted. Based on the dominant relation between different solutions, the individuals can be ranked as $F_0, F_1, ..., F_k$, where $F_0$ denotes the subset of individuals that have the overall best performance on both two metrics. Then, based on the ranking of individuals obtained, we determine whether to retain or remove them based on the following two rules. (1) An individual with a superior rank will have a higher probability of being reserved than the individual with an inferior rank. (2) If the two individuals are from the same rank, the one with a higher crowding distance will have a higher probability of being reserved. The details of the non-dominated sorting operation and crowding distance calculation can be found in the original NSGA-II strategy \cite{deb2002fast}.
	
	{\em \textbf{Step 6: Termination.}} If the total number of iterations exceeds the pre-defined maximum iterations, we terminate our method and output $F_0$ as Pareto fronts. Otherwise, we go back to {\em \textbf{Step 2}} and continue.

	\subsection{Biased Mutation} \label{sec:bias_mu}
	In the original mutation step of UCD, we simply determine the candidate node for disconnecting/connecting the corresponding neighbors/non-neighbors through a random selection, leading to suboptimal deception performance due to randomness. The above claim is also consistent with the corresponding results in Section \ref{sec:exp}. Thus, to further improve the attack performance, we propose the degree-biased and community-biased candidate node selection operations in the mutation process. First, as we aim to mislead the corresponding nodes to be classified into wrong clusters, it would be more powerful if we select the node pairs to be disconnected/connected with different levels of degrees, as the nodes with higher degree are typically situated in the core areas of the corresponding community while the nodes with lower degree are on the contrary. The above process can be considered as a degree-biased candidate node selection operation. Second, for the community-biased candidate node selection operation, based on the previous studies \cite{waniek2018hiding,chen2019ga,liu2022community}, the operation ``disconnect internal and connect external (DICE)'' tends to achieve a strong community hiding performance. In other words, the detection methods may be significantly influenced by disconnecting the nodes belonging to different communities and connecting the nodes belonging to the same community. 
	
	Based on the above analysis, we propose two variant strategies of our UCD by incorporating these two mechanisms into the mutation operation. Specifically, the details of the mutation operation ({\em \textbf{Step 3}}) of UCD will be improved as follows. First, we select the target node based on the degree of nodes. Thus, two specific variants can be obtained, namely UCD (MAX) and UCD (MIN). In UCD (MAX), nodes with a higher degree will have a higher probability of being selected as the target node. Under this scenario, nodes with a lower degree will have a higher probability of being selected as the nodes to be connected to/disconnected from this target node. In contrast, for UCD (MIN), the node selection preference is the opposite. At the same time, the DICE idea is also incorporated into the mutation operation. That is, the target nodes will prefer to connect a new link with the nodes belonging to different clusters and disconnect an existing link with the node belonging to the same cluster. Combining the two bias processes, we can summarize the following. For a target node having low/high degree, we will try to connect it with the nodes belonging to other clusters that have relatively high/low degree and disconnect it with the nodes belonging to the same cluster that has a relatively high/low degree. Under this situation, the target node can decrease its association with the current cluster and increase its association with other clusters as much as possible. Specifically, in our implementation, we utilize a roulette wheel selection to determine the selected target nodes based on their degree values. Aside from the above process, the other steps of UCD (MAX) and UCD (MIN) are the same as the original UCD method.

	
	\subsection{Overall Framework}
	The overall framework of the proposed community deception strategies is illustrated in Fig. \ref{fig:framework}. A series of individuals will be first generated through different strengths of perturbations in the \textit{Initialization} step. Then, \textit{Crossover} and \textit{Mutation} steps are utilized to combine the advantages of other individuals and induce new randomness. In particular, the biased mutation step is further proposed to improve the attack performance in UCD (MIN) and UCD (MAX). After that, \textit{Fitness Calculation} step is proposed to record the two considered metrics in this work, i.e., DARI and DAT. Finally, based on the quality of individuals with regard to the above metrics, the \textit{Elite Selection} step is used to determine the strong individuals to be reserved in the next iteration. The final perturbed graphs, which will have the corresponding unnoticeable perturbations injected from our strategies, can achieve a significant community deception performance under a limited attack resource.

	\section{Experiments}\label{sec:exp}

	\begin{table}[]
		\centering
		\renewcommand\arraystretch{1.1}
		\caption{Statistics of benchmark datasets.}
		\setlength{\tabcolsep}{20pt}
		\begin{tabular}{c||c|c}
			\bottomrule\bottomrule
			\textbf{Datesets} & \textbf{\# Nodes} & \textbf{\# Links}  \\ \hline\hline
			Karate              & 34           & 78                                         \\ \hline
			Dolphins          & 62           & 159                                                                                                       \\ \hline
			Netscience         & 379
			& 914      \\
			\bottomrule\bottomrule
		\end{tabular}
		\label{table:dataset}
	\end{table}

	\begin{figure*}[t]
		\subfigure[\textbf{Karate + LOU}]{
			\begin{minipage}[]{0.33\linewidth}
				\includegraphics[scale=0.33]{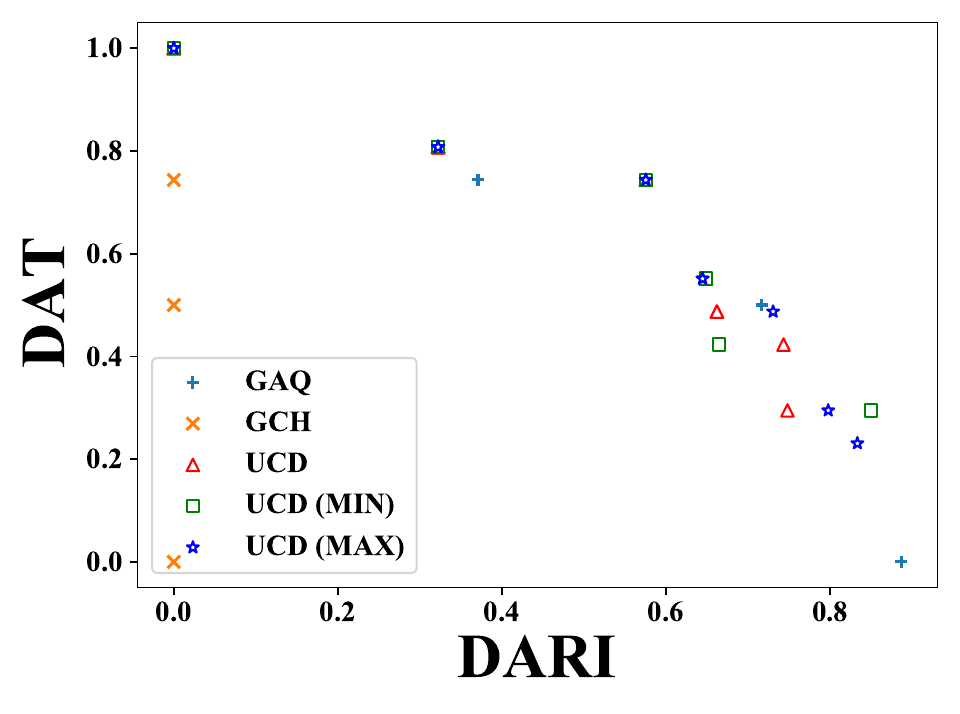}
			\end{minipage}%
		}%
		\subfigure[\textbf{Dolphins + LOU}]{
			\begin{minipage}[]{0.33\linewidth}
				\includegraphics[scale=0.33]{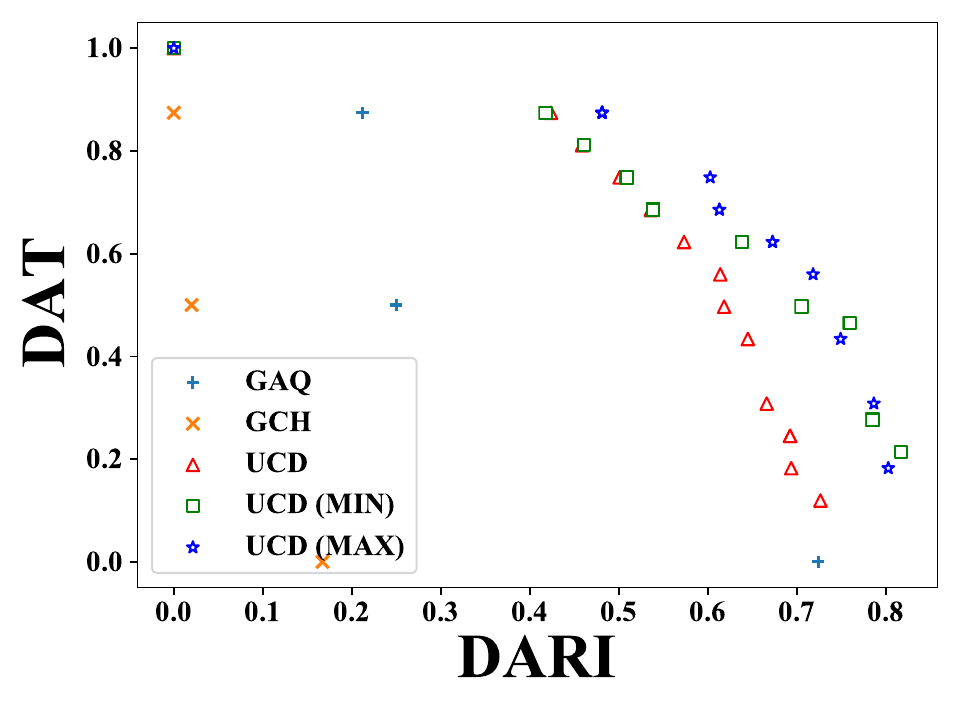}
			\end{minipage}%
		}%
		\subfigure[\textbf{Netscience + LOU}]{
			\begin{minipage}[]{0.33\linewidth}
				\includegraphics[scale=0.33]{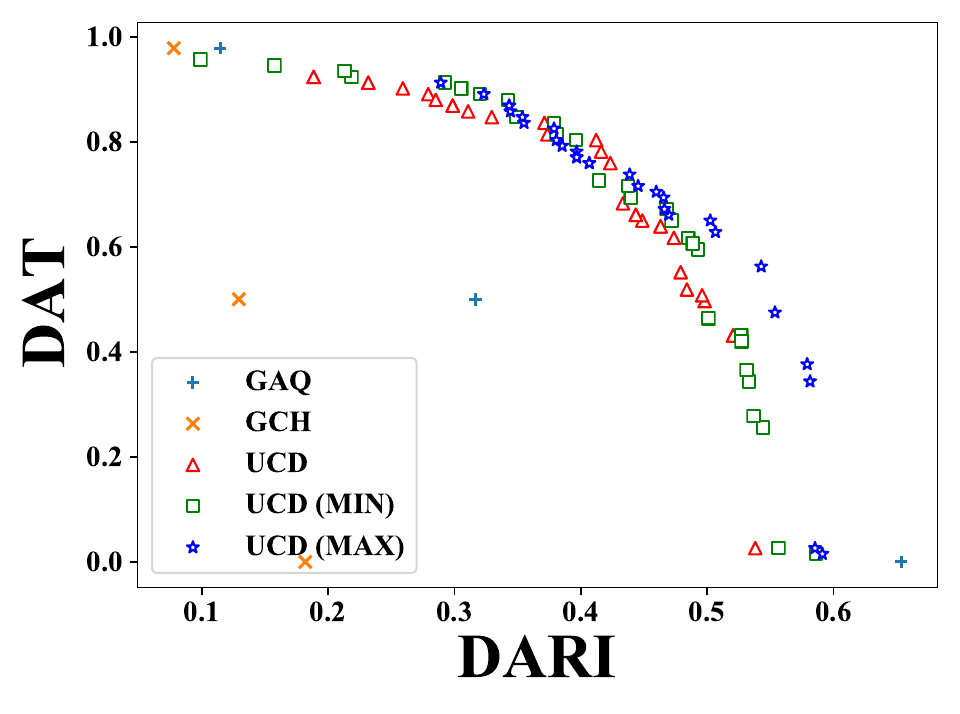}
			\end{minipage}%
		}%
		\\
		\subfigure[\textbf{Karate + FN}]{
			\begin{minipage}[]{0.33\linewidth}
				\includegraphics[scale=0.33]{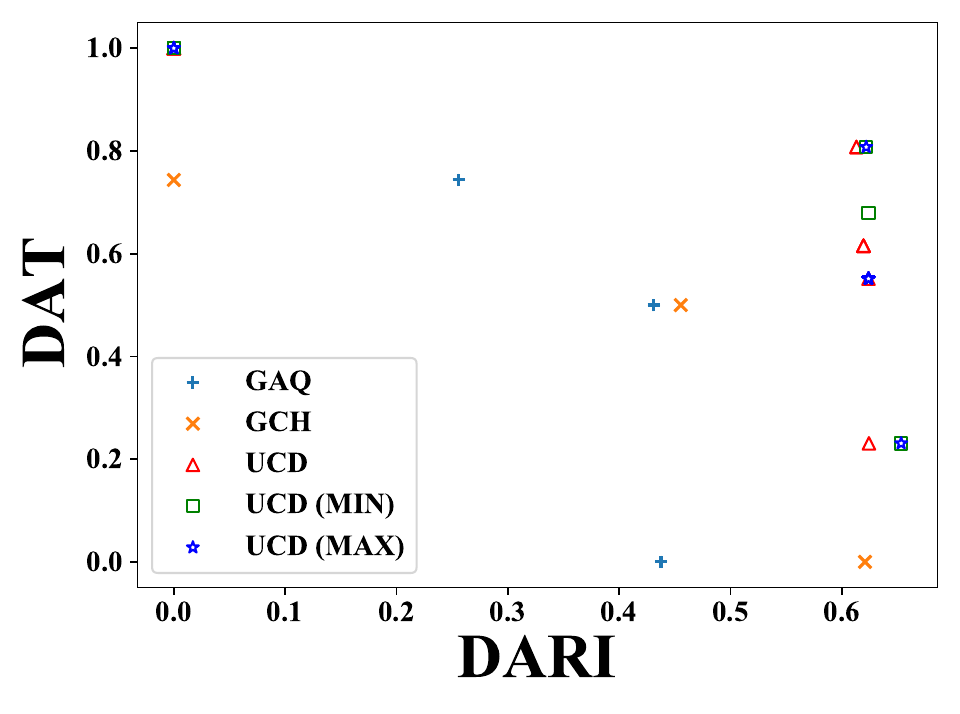}
			\end{minipage}%
		}%
		\subfigure[\textbf{Dolphins + FN}]{
			\begin{minipage}[]{0.33\linewidth}
				\includegraphics[scale=0.33]{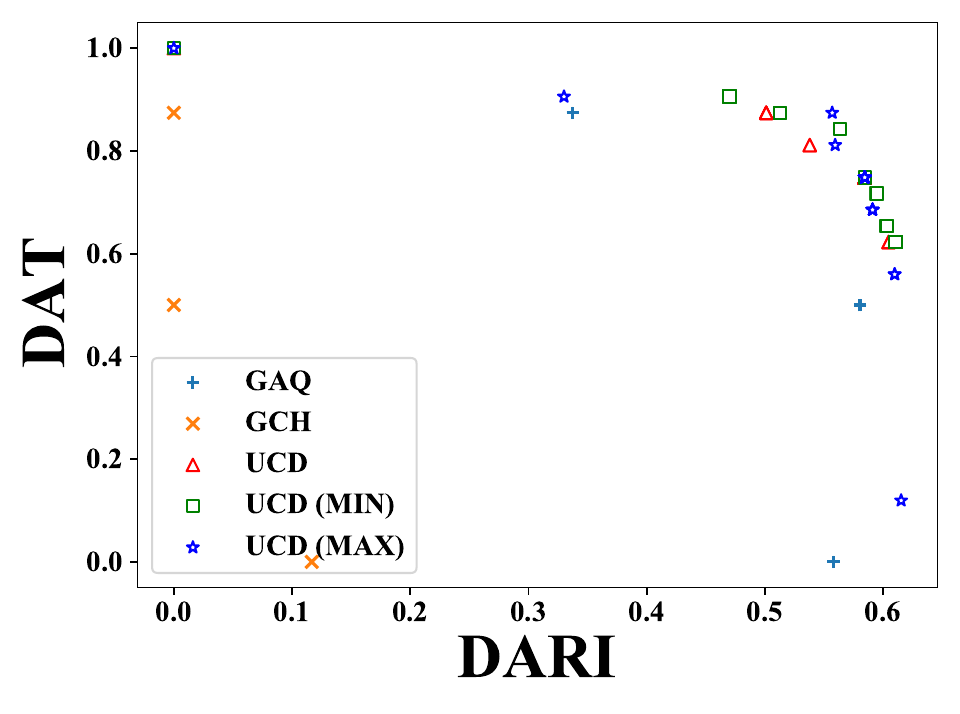}
			\end{minipage}%
		}%
		\subfigure[\textbf{Netscience +FN}]{
			\begin{minipage}[]{0.33\linewidth}
				\includegraphics[scale=0.33]{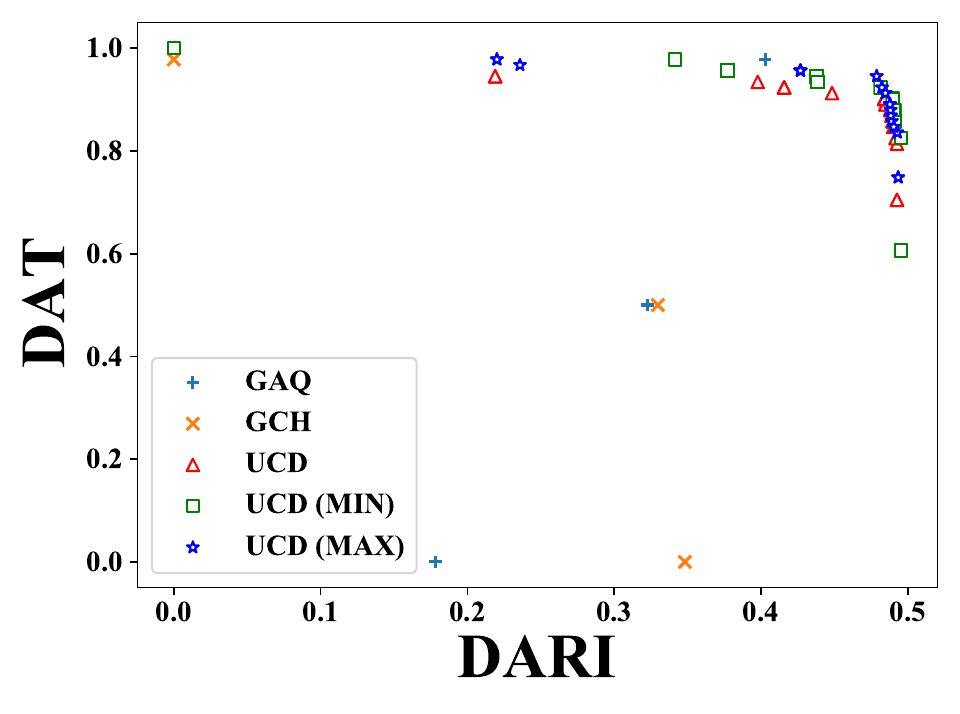}
			\end{minipage}%
		}%
		\\
		\subfigure[\textbf{Karate + LPA}]{
			\begin{minipage}[]{0.33\linewidth}
				\includegraphics[scale=0.33]{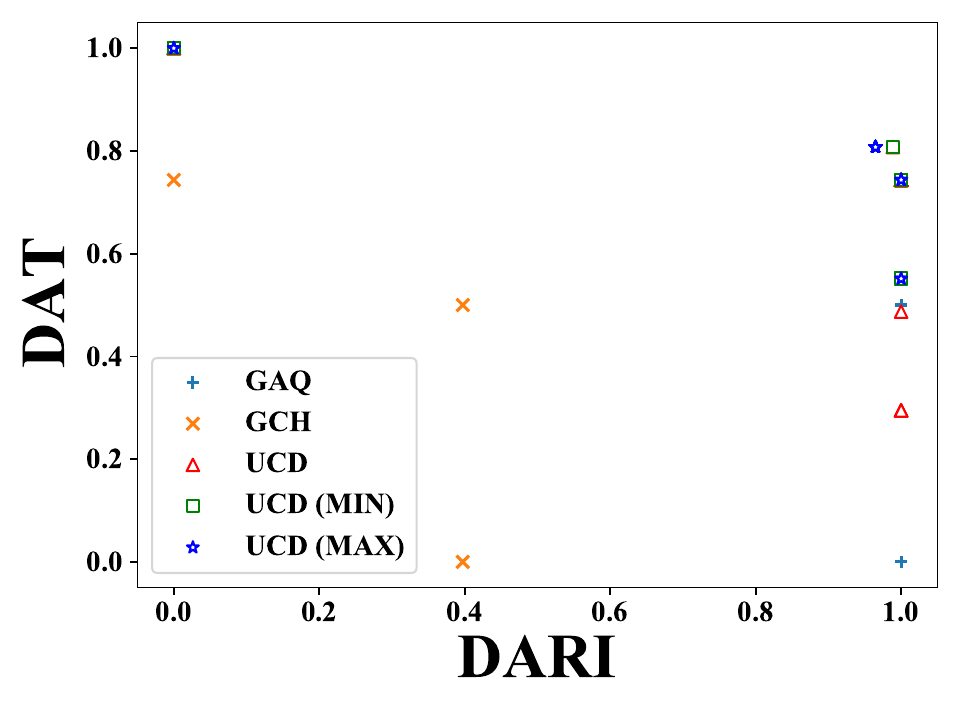}
			\end{minipage}%
		}%
		\subfigure[\textbf{Dolphins + LPA}]{
			\begin{minipage}[]{0.33\linewidth}
				\includegraphics[scale=0.33]{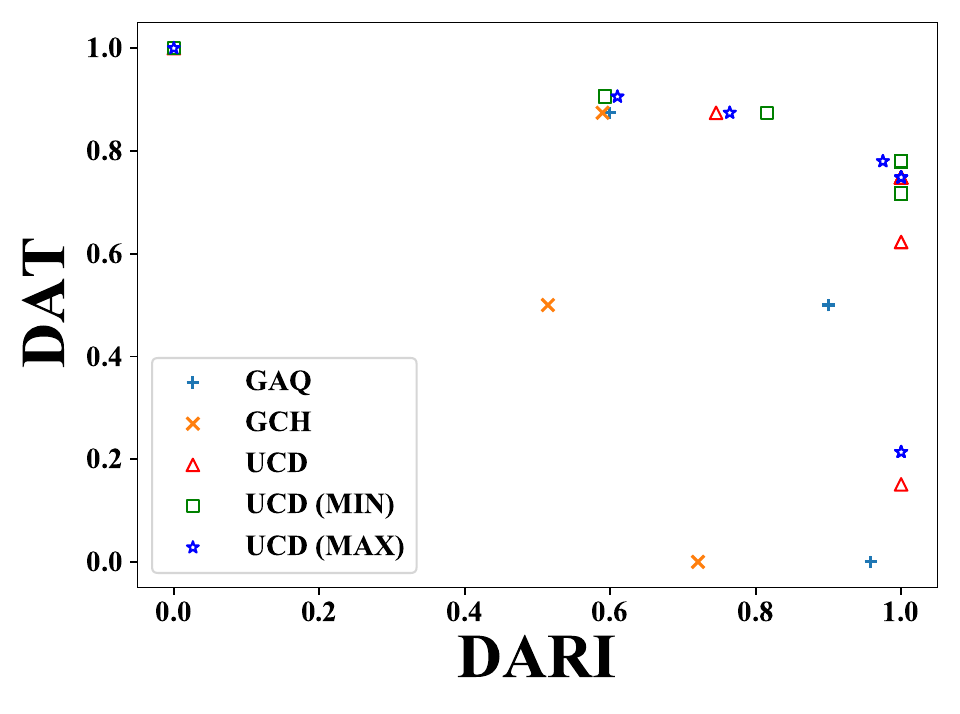}
			\end{minipage}%
		}%
		\subfigure[\textbf{Netscience + LPA}]{
			\begin{minipage}[]{0.33\linewidth}
				\includegraphics[scale=0.33]{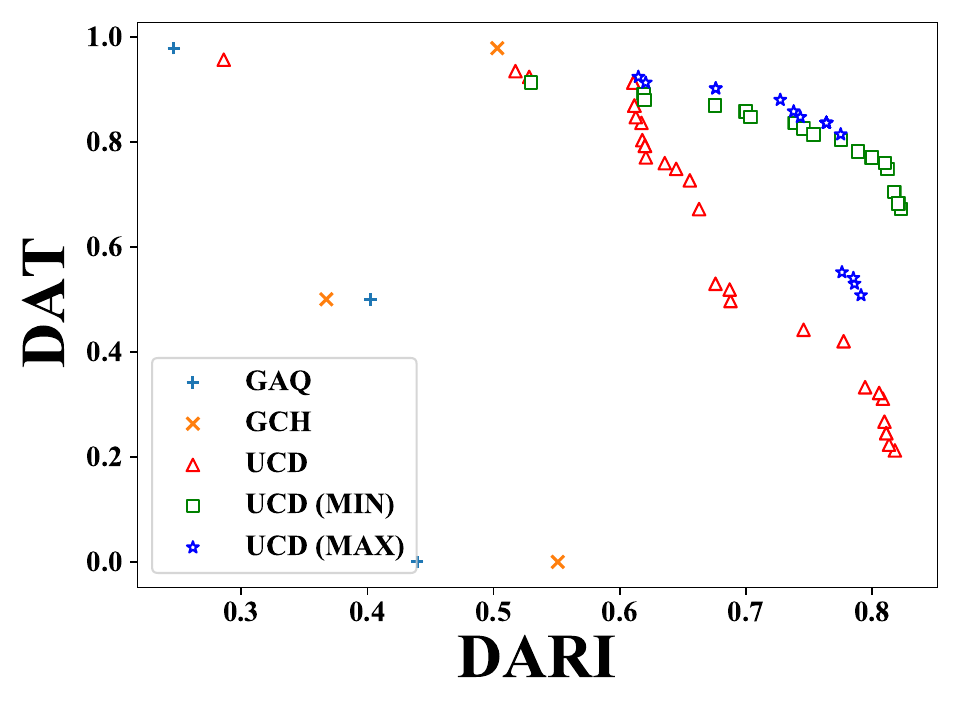}
			\end{minipage}%
		}%

		\centering
		\caption{Comparison of the Pareto fronts of UCDs and baseline methods under three detection algorithms on three datasets. The red triangles, green squares, and blue stars indicate the solutions of Pareto fronts obtained by UCD, UCD (MIN), and UCD (MAX), respectively. The sky blue plus and orange cross denote the representative solutions of GAQ and GCH, respectively.}
		\label{fig:pf}
	\end{figure*}

	In this section, we conduct comprehensive experiments to verify the effectiveness of the proposed UCDs. The tested datasets, selected detection methods, baselines, and parameter settings are introduced first. Then, the detailed experimental results along with corresponding discussions are presented.

	\subsection{Datasets}
	In the experiments, three classic datasets that are widely used for the community detection tasks are chosen. The details of these datasets are as follows.
	
  \begin{enumerate}
		
		\item{\textbf{Karate \cite{zachary1977information}.}} The Karate dataset is a social network built by observing a karate club at an American university. The nodes in the network represent the members of the club and the links represent friendship relationships among members.
		
		\item{\textbf{Dolphins \cite{lusseau2003bottlenose}.}} The dolphin dataset is built by observing the living habits of 62 bottlenose dolphins in New Zealand. Considering each dolphin as a node, a link exists between two nodes if their corresponding two dolphins frequently move together.
		
		
		\item{\textbf{Netscience \cite{newman2006finding}.}} The NetScience dataset is about the collaborative relationships among co-authors of network science. Each node represents a scientist and each link between two nodes represents a collaborative relationship between two scientists.
	\end{enumerate}
	The detailed statistics for the three aforementioned datasets are given in Table \ref{table:dataset}.

	\begin{figure}[]
		\centering
		
		\includegraphics[width=0.5\linewidth]{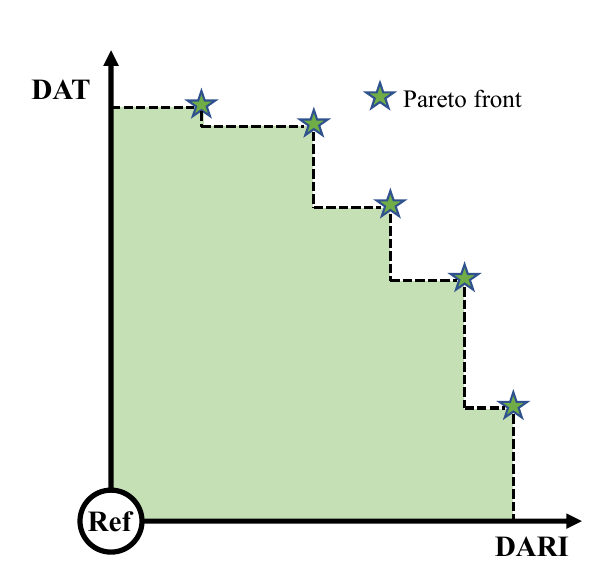} 
		
		\caption{Illustration of the hyper volume with (0, 0) as the reference point.}
		\label{fig:hv_vis}
	\end{figure}

	\begin{figure*}[t]
		\subfigure[\textbf{Karate + LOU}]{
			\begin{minipage}[]{0.33\linewidth}
				\includegraphics[scale=0.33]{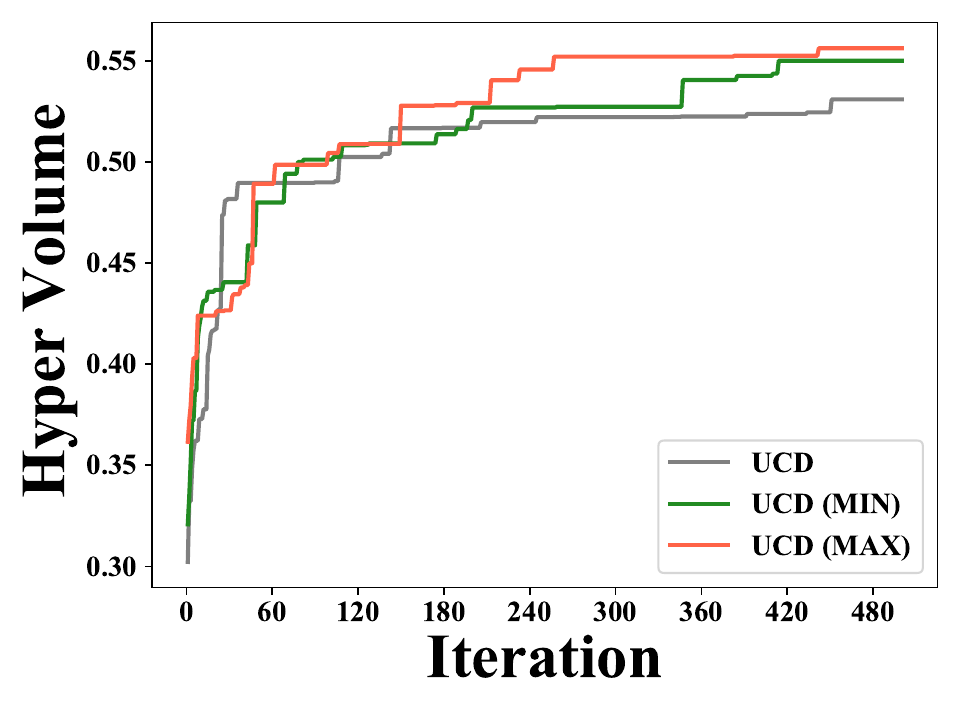}
			\end{minipage}%
		}%
		\subfigure[\textbf{Dolphins + LOU}]{
			\begin{minipage}[]{0.33\linewidth}
				\includegraphics[scale=0.33]{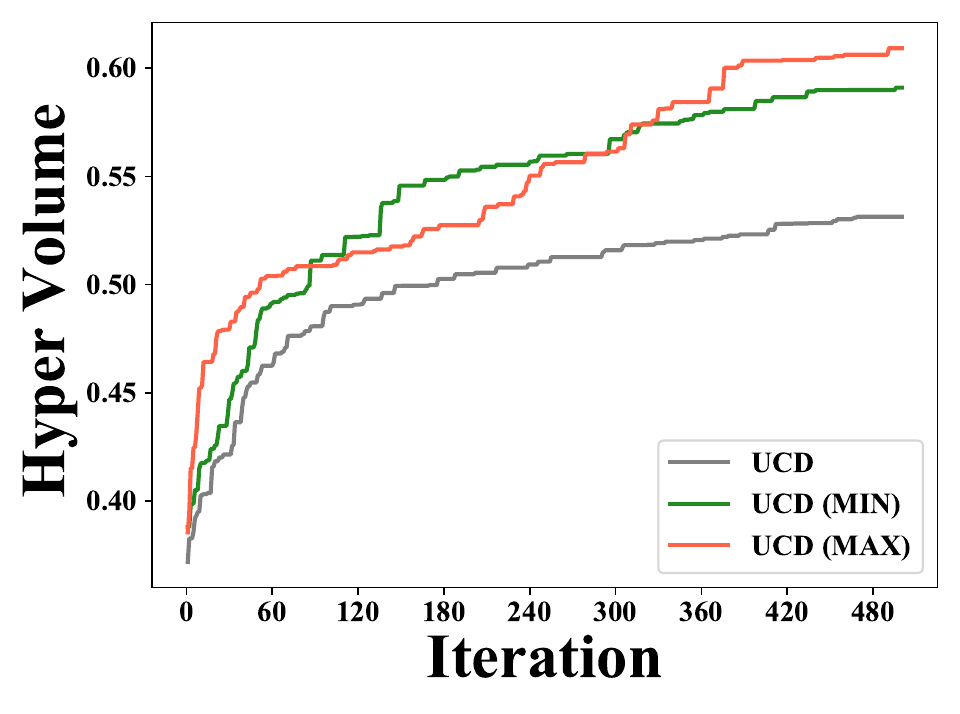}
			\end{minipage}%
		}%
		\subfigure[\textbf{Netscience + LOU}]{
			\begin{minipage}[]{0.33\linewidth}
				\includegraphics[scale=0.33]{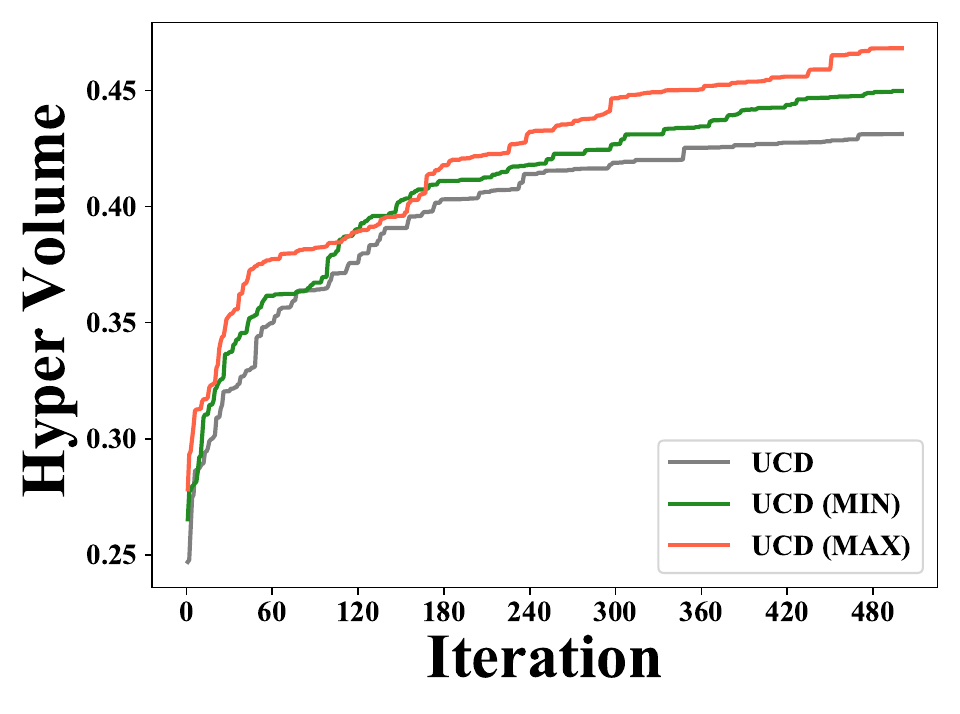}
			\end{minipage}%
		}%
		\\
		\subfigure[\textbf{Karate + FN}]{
			\begin{minipage}[]{0.33\linewidth}
				\includegraphics[scale=0.33]{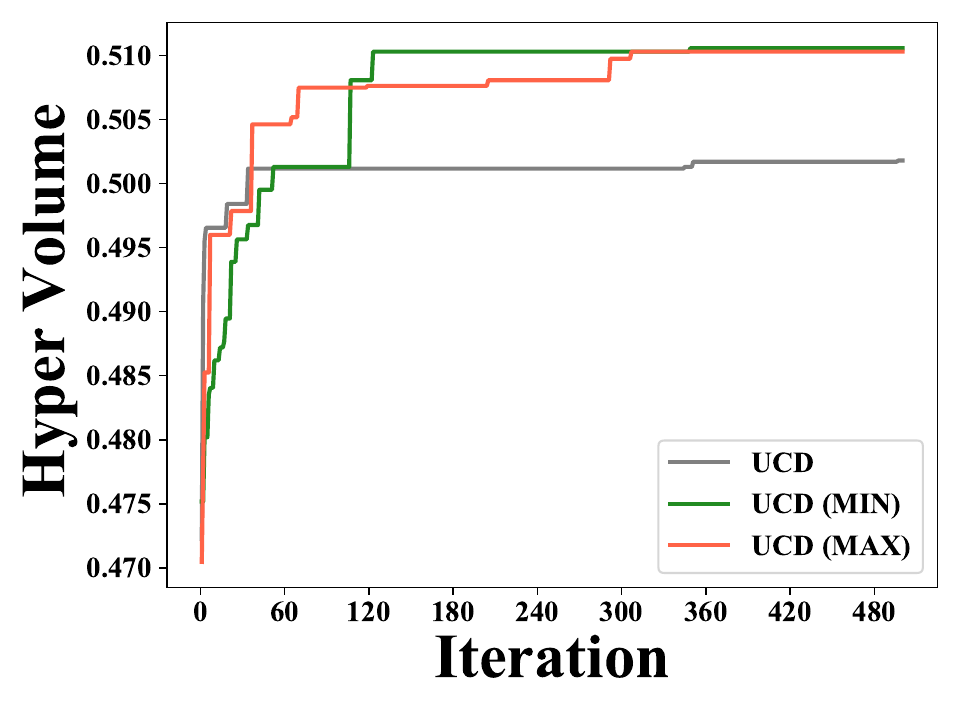}
			\end{minipage}%
		}%
		\subfigure[\textbf{Dolphins + FN}]{
			\begin{minipage}[]{0.33\linewidth}
				\includegraphics[scale=0.33]{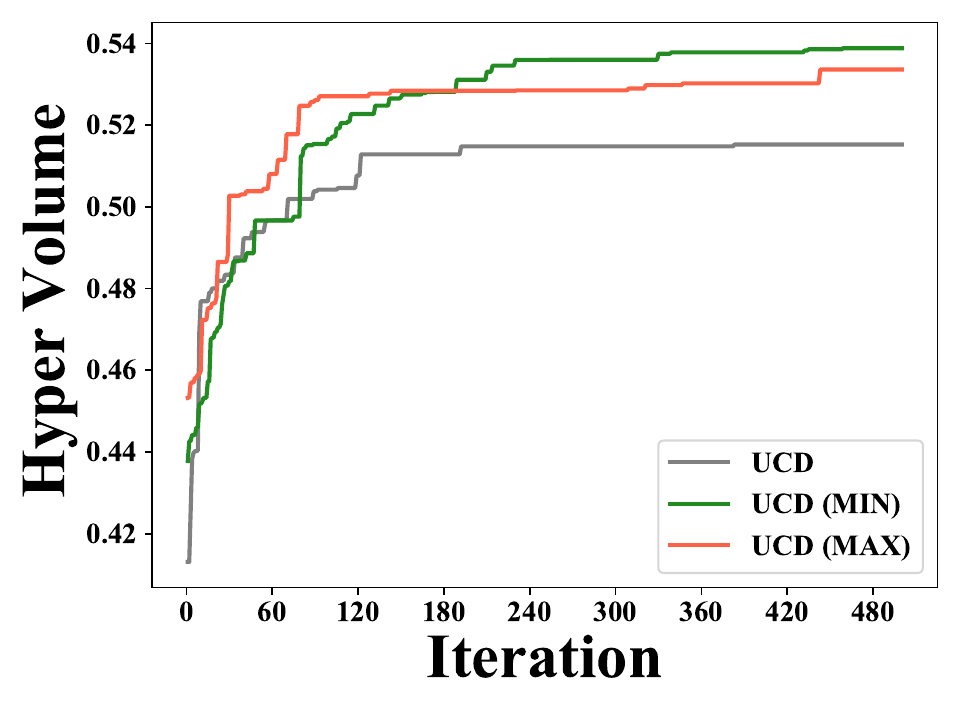}
			\end{minipage}%
		}%
		\subfigure[\textbf{Netscience + FN}]{
			\begin{minipage}[]{0.33\linewidth}
				\includegraphics[scale=0.33]{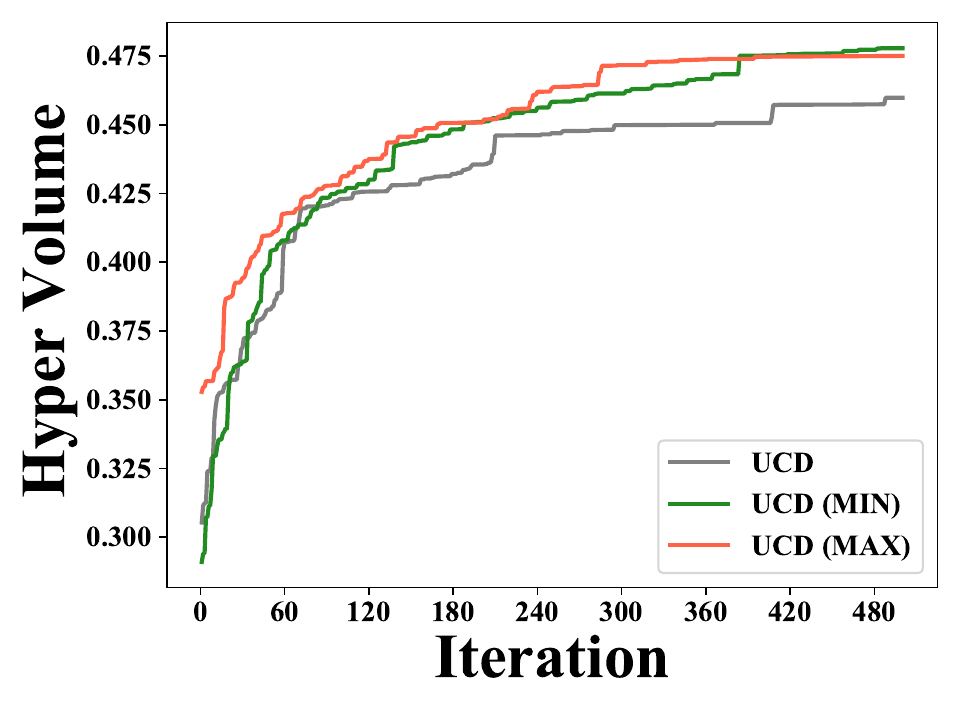}
			\end{minipage}%
		}%
		\\
		\subfigure[\textbf{Karate + LPA}]{
			\begin{minipage}[]{0.33\linewidth}
				\includegraphics[scale=0.33]{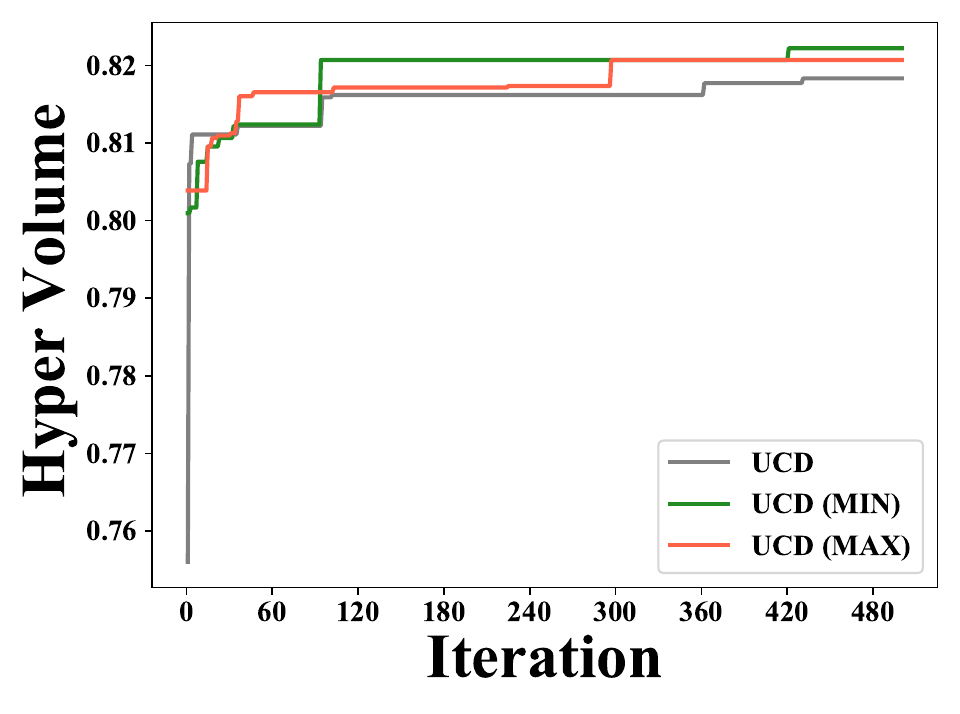}
			\end{minipage}%
		}%
		\subfigure[\textbf{Dolphins + LPA}]{
			\begin{minipage}[]{0.33\linewidth}
				\includegraphics[scale=0.33]{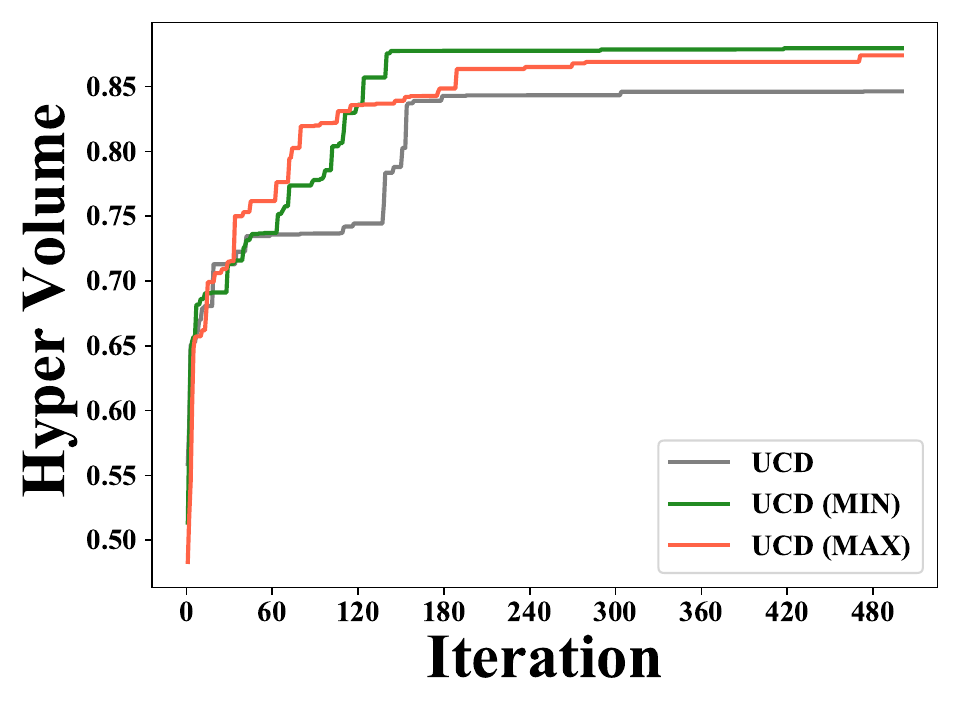}
			\end{minipage}%
		}%
		\subfigure[\textbf{Netscience + LPA}]{
			\begin{minipage}[]{0.33\linewidth}
				\includegraphics[scale=0.33]{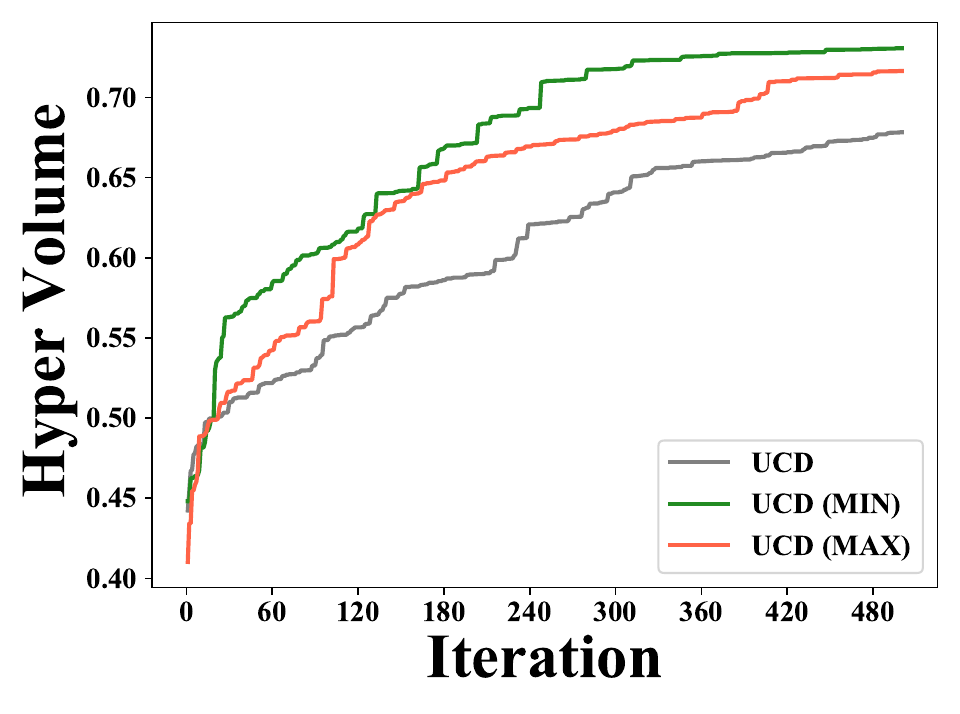}
			\end{minipage}%
		}%
		\centering
		\caption{Comparison of the changes in hyper volume of UCDs with regard to the number of iterations under three detection algorithms on three datasets.}
		\label{fig:hv}
	\end{figure*}

	\begin{table*}[]
		\centering
		\caption{Comparison of the average hyper volume and solution diversity of UCDs under three detection algorithms on three datasets.}
		\begin{tabular}{c||c|cc|cc|cc}
			\bottomrule\bottomrule
			\multirow{2}{*}{\textbf{Datasets}} & \multirow{2}{*}{\textbf{Detection Algorithms}} & \multicolumn{2}{c|}{\textbf{UCD}} & \multicolumn{2}{c|}{\textbf{UCD (MIN)}} & \multicolumn{2}{c}{\textbf{UCD (MAX)}} \\ \cline{3-8} 
			& & \multicolumn{1}{c|}{\textbf{Hyper Volume}} & \textbf{Diversity} & \multicolumn{1}{c|}{\textbf{Hyper Volume}} & \textbf{Diversity} & \multicolumn{1}{c|}{\textbf{Hyper Volume}} & \textbf{Diversity} \\ \hline\hline
			\multirow{3}{*}{Karate} 
			& LOU & \multicolumn{1}{c|}{0.5429} & 14.76 & \multicolumn{1}{c|}{0.5454} & 15.40 & \multicolumn{1}{c|}{\textbf{0.5478}} & \textbf{15.56} \\ \cline{2-8}
			& FN & \multicolumn{1}{c|}{0.5046} & 16.98 & \multicolumn{1}{c|}{\textbf{0.5047}} & 16.56 & \multicolumn{1}{c|}{0.5042} & \textbf{17.34} \\ \cline{2-8}
			& LPA & \multicolumn{1}{c|}{\textbf{0.8215}} & 25.10 & \multicolumn{1}{c|}{0.8213} & 24.32 & \multicolumn{1}{c|}{0.8212} & \textbf{25.36} \\ \hline
			
			\multirow{3}{*}{Dolphins} 
			& LOU & \multicolumn{1}{c|}{0.5612} & 24.18 & \multicolumn{1}{c|}{\textbf{0.5642}} & 22.84 & \multicolumn{1}{c|}{0.5639} & \textbf{25.10} \\ \cline{2-8}
			& FN & \multicolumn{1}{c|}{0.5228} & 22.28 & \multicolumn{1}{c|}{0.5235} & \textbf{24.04} & \multicolumn{1}{c|}{\textbf{0.5238}} & 22.98 \\ \cline{2-8}
			& LPA & \multicolumn{1}{c|}{0.8529} & 26.36 & \multicolumn{1}{c|}{0.8575} & \textbf{27.04} & \multicolumn{1}{c|}{\textbf{0.8577}} & 25.32 \\ \hline
			
			\multirow{3}{*}{Netscience} 
			& LOU & \multicolumn{1}{c|}{0.4367} & 28.80 & \multicolumn{1}{c|}{\textbf{0.4385}} & 28.98 & \multicolumn{1}{c|}{0.4380} & \textbf{29.08} \\ \cline{2-8}
			& FN & \multicolumn{1}{c|}{0.4574} & 29.82 & \multicolumn{1}{c|}{\textbf{0.4594}} & \textbf{29.44} & \multicolumn{1}{c|}{0.4581} & 28.74 \\ \cline{2-8}
			& LPA & \multicolumn{1}{c|}{0.6696} & 25.66 & \multicolumn{1}{c|}{\textbf{0.6771}} & 26.40 & \multicolumn{1}{c|}{0.6719} & \textbf{27.34} \\ \bottomrule\bottomrule
		\end{tabular}
		\label{tab:dac}
	\end{table*}

	\subsection{Evaluated Detection Algorithms}
	In the experiments, three classic detection methods, including LOU, FN, and LPA, are selected as the targeted algorithms to be attacked. The details of them are given as follows.  
	  \begin{enumerate}
		\item \textbf{LOU (Louvain Algorithm) \cite{blondel2008fast}} LOU is a classic and powerful community detection method that identifies community structures by maximizing modularity. It initially considers each node as an independent community and then iteratively merges nodes to optimize modularity. The above process continues until modularity no longer increases.
		\item \textbf{FN (Fast Newman Algorithm) \cite{newman2004fast}} FN is a community detection method that optimizes modularity based on a greedy strategy, which starts with each node as an independent community and iteratively merges communities to maximize modularity. Compared to the LOU method, FN enhances efficiency through heuristic community merging. Ultimately, FN selects the community partition with the highest modularity as the final output.
		\item \textbf{LPA (Label Propagation Algorithm) \cite{raghavan2007near}} LPA identifies communities by propagating labels through the network. At the start, each node in the network is assigned a unique label. During subsequent iterations, each node updates its label based on the labels of its neighbors, continuing until the labels no longer change.
	\end{enumerate}

	\subsection{Baselines}
	To comprehensively demonstrate the effectiveness of the proposed method, two state-of-the-art deception methods are selected as baselines, as follows.

  \begin{enumerate}
		\item \textbf{GAQ \cite{chen2019ga}} (GA-based Q attack) GAQ employs the genetic algorithm to obtain the optimal attack perturbations, where the modularity of networks is used to design the fitness function.
		\item \textbf{GCH \cite{liu2022community}} (Graph community hiding using a graph auto-encoder) GCH is a novel method that adopts the graph auto-encoder \cite{kipf2016variational} to achieve community hiding. First, a graph auto-encoder comprising an encoder and a decoder is trained to create a probability adjacency matrix. Then, based on the guidance of the probability matrix and a given threshold, GCH deletes the corresponding links in the same community and adds the corresponding links in different communities. To extend this method into a more scalable version, instead of pre-defining a threshold, we will greedily select the corresponding perturbed links based on the prediction probabilities. 
	\end{enumerate}

	\subsection{Experimental Setups}
	For the proposed UCDs, the size of population $\Omega$ is set to 30. The crossover rate and mutation rate of UCDs are set to 0.5 and 0.8, respectively. The maximal budget of link modification is set to 20\% of the number of original links. The maximal iteration of UCDs is set to 500. 
	
	Moreover, for the GAQ method, we set all the parameters to be the same as the proposed UCDs to ensure a fair comparison. For the GCH method, we set the encoder to be a two-layer GCN \cite{kipf2017semi}. It's worth noting that both of the above two methods fail to guarantee the degree distribution of networks before and after the attacks.

	\begin{figure*}[]
		\subfigure[\textbf{Original network}]{
			\begin{minipage}[]{0.33\linewidth}
				\includegraphics[scale=0.35]{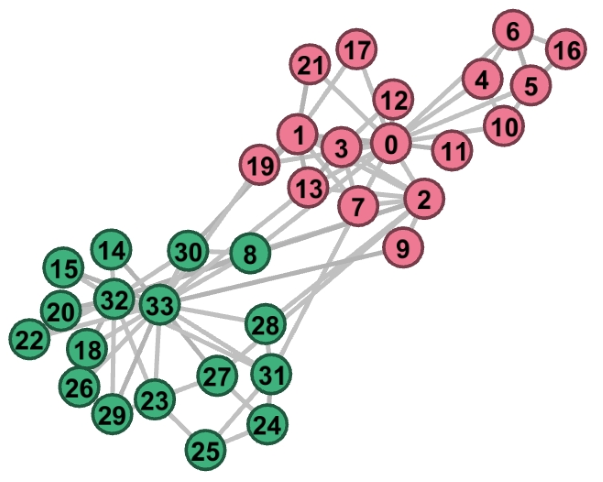}
			\end{minipage}%
		}%
		\subfigure[\textbf{GAQ}]{
			\begin{minipage}[]{0.33\linewidth}
				\includegraphics[scale=0.35]{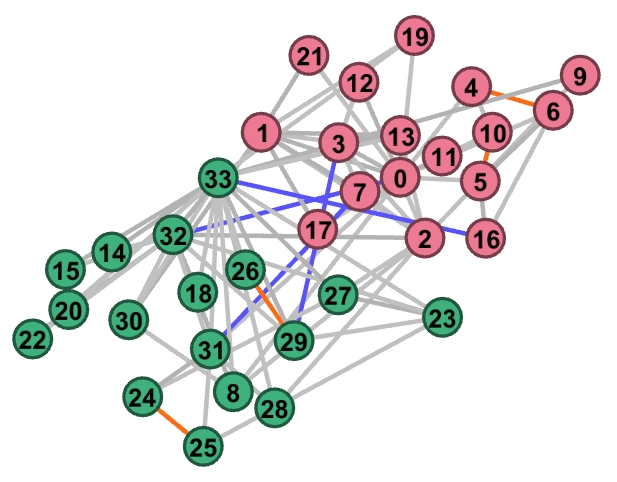}
			\end{minipage}%
		}%
		\subfigure[\textbf{GCH}]{
			\begin{minipage}[]{0.33\linewidth}
				\includegraphics[scale=0.35]{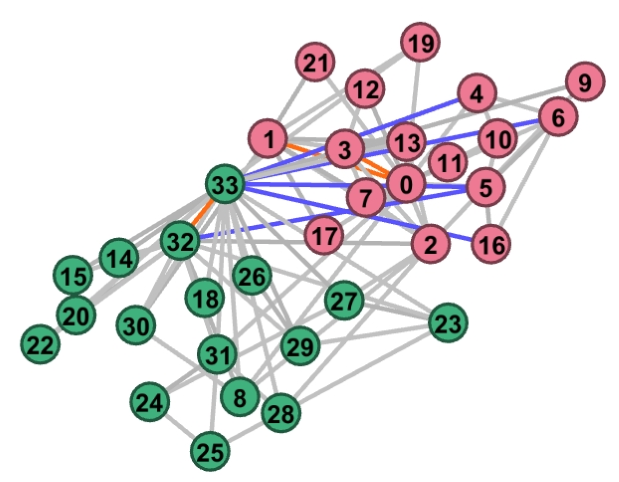}
			\end{minipage}%
		}%
		\\
		\subfigure[\textbf{UCD}]{
			\begin{minipage}[]{0.33\linewidth}
				\includegraphics[scale=0.35]{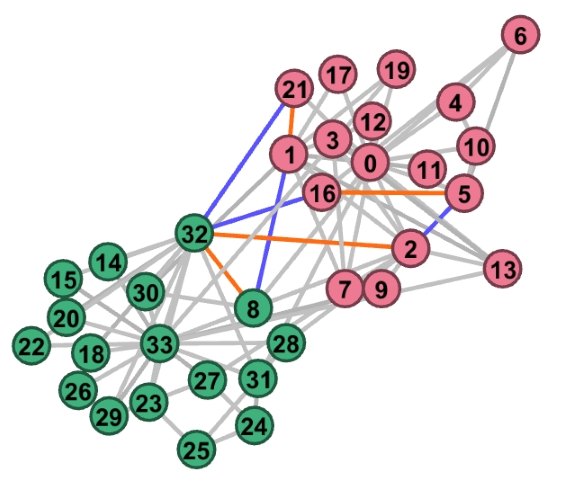}
			\end{minipage}%
		}%
		\subfigure[\textbf{UCD (MIN)}]{
			\begin{minipage}[]{0.33\linewidth}
				\includegraphics[scale=0.35]{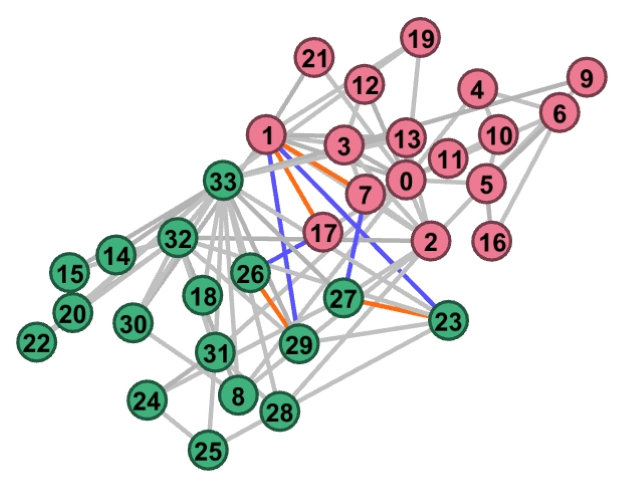}
			\end{minipage}%
		}%
		\subfigure[\textbf{UCD (MAX)}]{
			\begin{minipage}[]{0.33\linewidth}
				\includegraphics[scale=0.32]{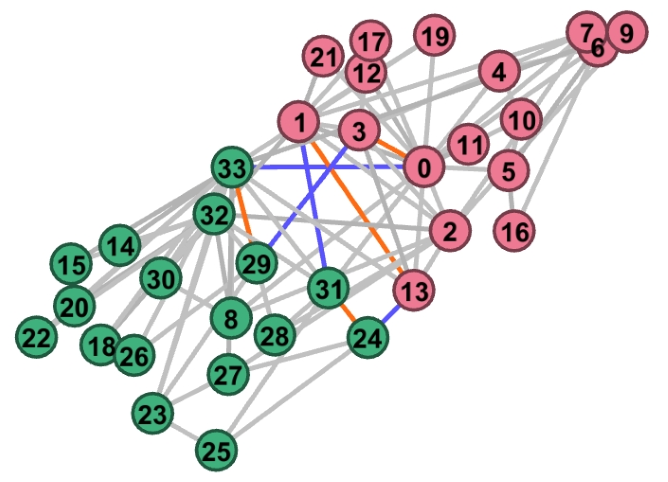}
			\end{minipage}%
		}%
		\centering
		\caption{Visualization of the perturbed karate networks generated by corresponding community deception strategies. The green and red nodes denote the nodes belonging to two different clusters (obtained by the LOU method), respectively. The blue and orange lines represent the newly added and deleted malicious links, respectively.}
		\label{fig:vis}
	\end{figure*}

	\subsection{Results}
	\subsubsection{Pareto Fronts}
	 To evaluate the effectiveness of our proposed methods, we first demonstrate their Pareto fronts, a widely used measure in multi-objective optimizations. The Pareto fronts illustrate the trade-off between different objectives in multi-objective optimization, providing a comprehensive perspective to characterize the performance of corresponding multi-objective optimization strategies. Specifically, in this work, the two objectives we focus on are the decrease of adjusted rand index (DARI) and the decrease of attack budget (DAT), and our goal is to maximize the above two metrics, aiming to achieve a more powerful community deception performance under a smaller budget constraint. Moreover, as GAQ and GCH are not directly designed for the multi-objective optimization scenario, we manually select three representative solutions, namely the farthest left one, exact center one, and farthest right one, by setting the corresponding attack budget to compare with the Pareto fronts obtained by the proposed UCDs.
	 
	 Fig. \ref{fig:pf} demonstrates the obtained Pareto fronts of our strategies and representative solutions of two baseline methods. As we can observe, three UCDs can obtain a series of optimal solutions across a different range of metrics. For the two single-objective baselines, GAQ shows a much better performance than GCH in all datasets. For the comparison between multi- and single-objective methods, we find that the proposed UCDs can obtain comparable or even better solutions than GAQ and GCH, indicating the superiority of multi-objective optimizations. In addition, for the comparison of the proposed three UCDs, the two variant methods combining the biased mutation process exhibit an overall better performance than the original UCD, which also indicates the positive effect of degree-biased and community-biased candidate node selection mechanisms.

	\subsubsection{Hyper Volume and Solution Diversity}
	Besides the qualitative analysis of the proposed UCDs, we also quantitatively investigate their performance. Hyper volume is a widely used metric to evaluate the performance of multi-objective optimization algorithms. A simple illustration of hyper volume, which can be calculated through the area of the shaded region between the Pareto fronts and a standard reference point (i.e., (0, 0) in this work),  is given in Fig. \ref{fig:hv_vis}. As we focus on maximizing the considered two metrics, i.e., DARI and DAT, we prefer the area of the shaded region to be as large as possible. In addition to the hyper volume, we are also curious about the diversity of the obtained Pareto fronts, i.e., the number of solutions, where more diverse solutions indicate stronger exploration abilities of the corresponding strategy.
	
	The detailed results of the obtained hyper volume and diversity of the three UCDs are given in Table \ref{tab:dac}. We can observe that all three methods tend to obtain the Pareto fronts with pretty good diversity, especially for the two variant methods UCD (MIN) and UCD (MAX). For the hyper volume, we further investigate it from a micro perspective by presenting the change in hyper volume with regard to the running iteration on three datasets, as given in Fig. \ref{fig:hv}. We can find that the two variant methods tend to obtain a higher hyper volume than the original UCD almost all the time. The above results demonstrate the deception effect and comprehensiveness of the proposed UCDs, especially for the two variant methods.

	\subsubsection{Visualization of Perturbed Networks}
	Then, aiming to deeply investigate the hiding effect of the proposed UCDs, we visualize the corresponding perturbed networks generated by them. Fig. \ref{fig:vis} demonstrates the perturbed networks of baselines and proposed UCDs. For the attack patterns, as can be observed, GAQ follows the DICE-like attack patterns, while GCH does not. The above results also support the unsatisfactory deception performance of GCH in Fig. \ref{fig:pf}. Moreover, the proposed three UCDs mainly follow the DICE attack pattern, except for the original UCD (i.e., newly added link $2-5$). In particular, the degree-biased pattern is also observed in these two variant methods, which is consistent with our designs in Section \ref{sec:bias_mu}. 
	
	In addition, in terms of the unnoticeability, we can find that both GAQ and GCH violate the constraints of degree distributions, as they simply added or removed the corresponding links while ignoring the local degree of those nodes. Instead, the proposed UCDs can make sure the degree of each node remains the same before and after the attacks, which improves the unnoticeability of the community deception to some extent.

	\begin{figure*}[]
		\subfigure[\textbf{Karate + LOU}]{
			\begin{minipage}[]{0.33\linewidth}
				\includegraphics[scale=0.33]{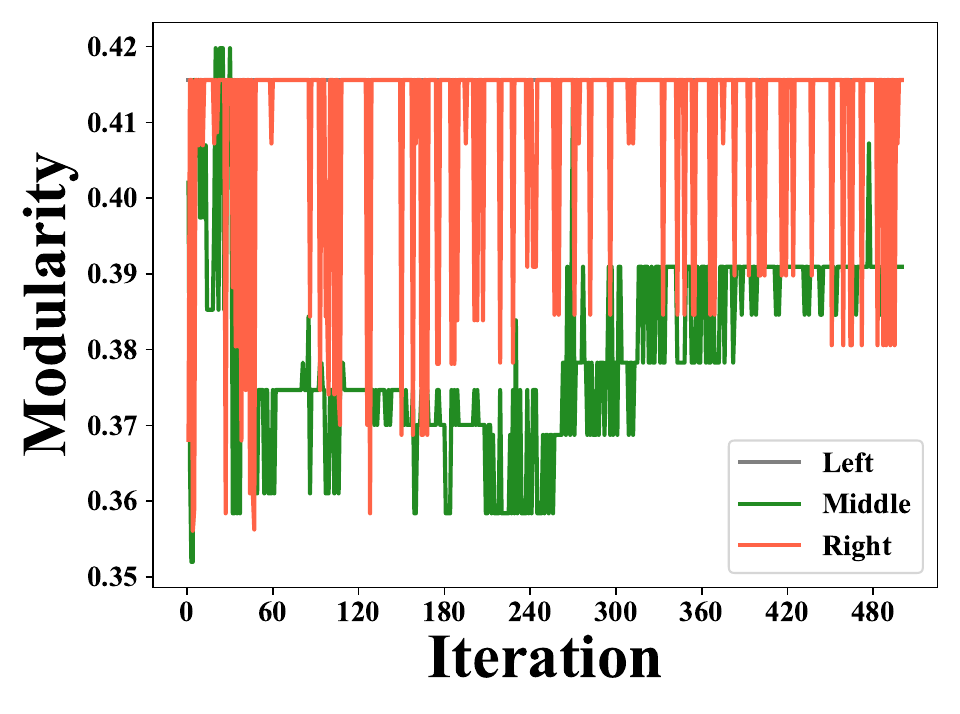}
			\end{minipage}%
		}%
		\subfigure[\textbf{Dolphins + LOU}]{
			\begin{minipage}[]{0.33\linewidth}
				\includegraphics[scale=0.33]{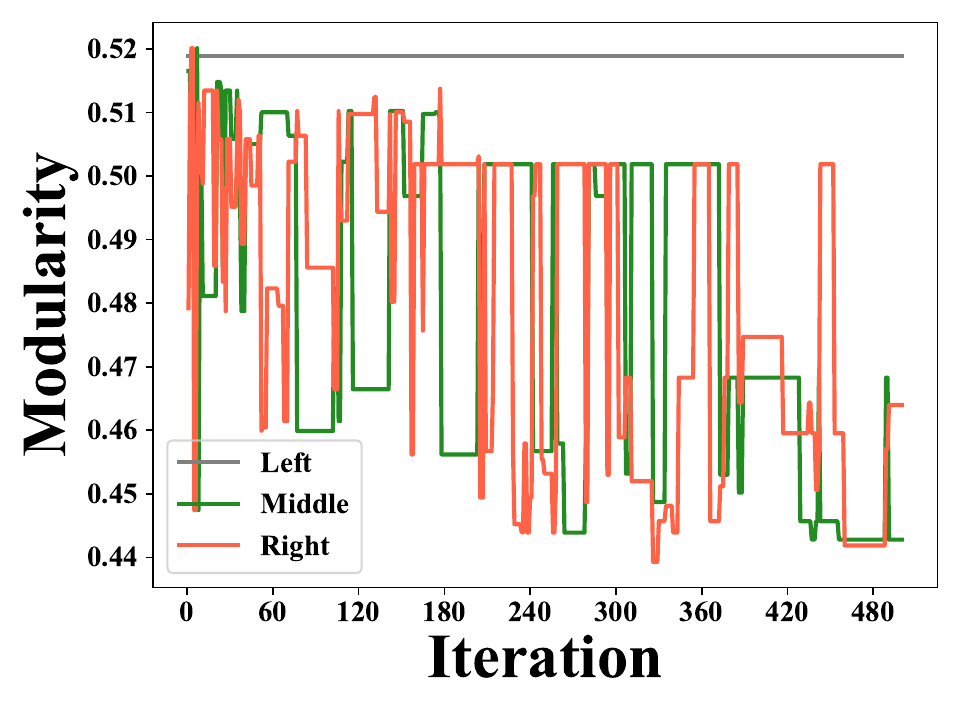}
			\end{minipage}%
		}%
		\subfigure[\textbf{Netscience + LOU}]{
			\begin{minipage}[]{0.33\linewidth}
				\includegraphics[scale=0.33]{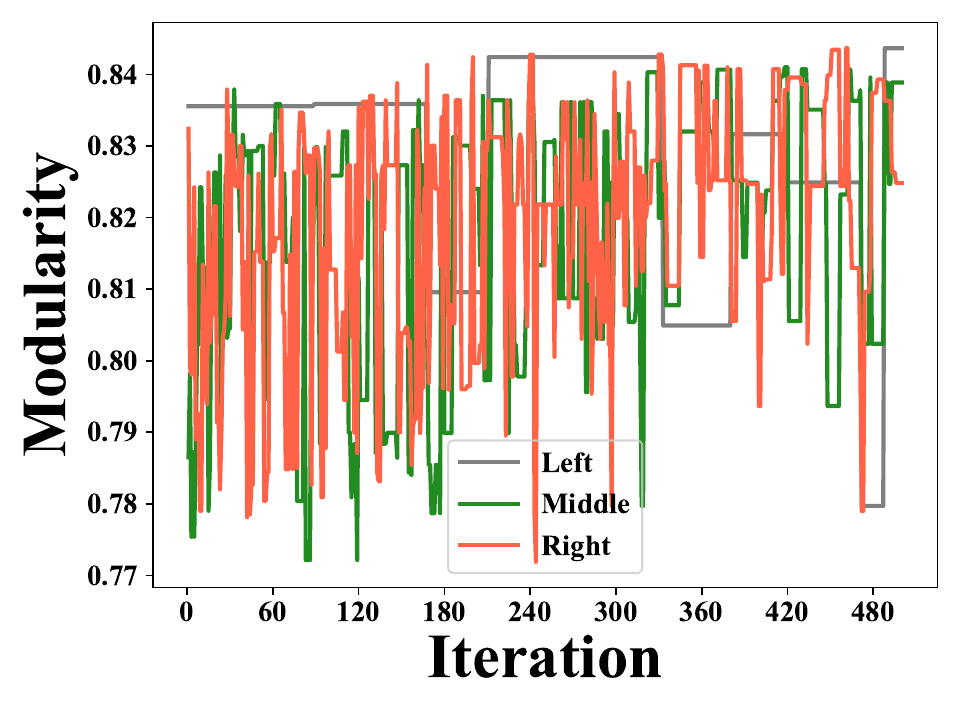}
			\end{minipage}%
		}%
		\\
		\subfigure[\textbf{Karate + FN}]{
			\begin{minipage}[]{0.33\linewidth}
				\includegraphics[scale=0.33]{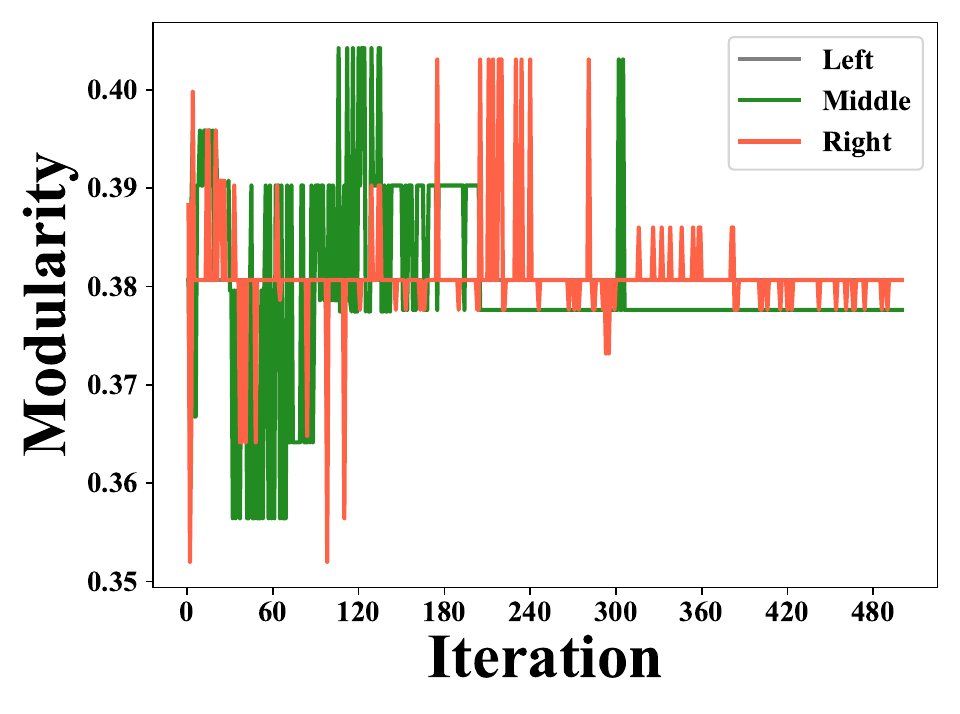}
			\end{minipage}%
		}%
		\subfigure[\textbf{Dolphins + FN}]{
			\begin{minipage}[]{0.33\linewidth}
				\includegraphics[scale=0.33]{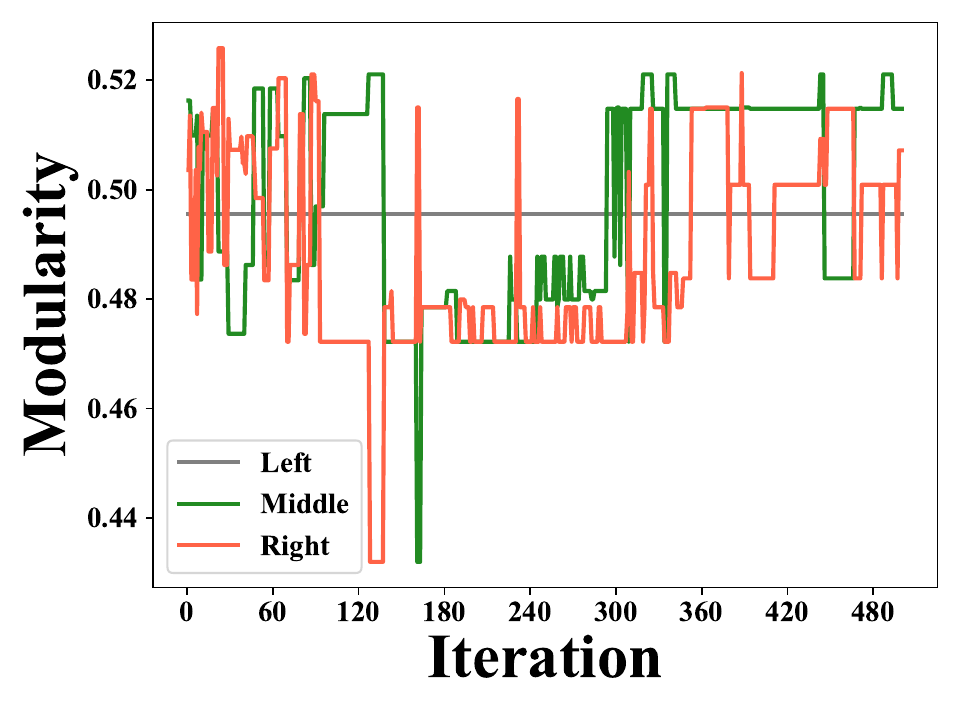}
			\end{minipage}%
		}%
		\subfigure[\textbf{Netscience + FN}]{
			\begin{minipage}[]{0.33\linewidth}
				\includegraphics[scale=0.33]{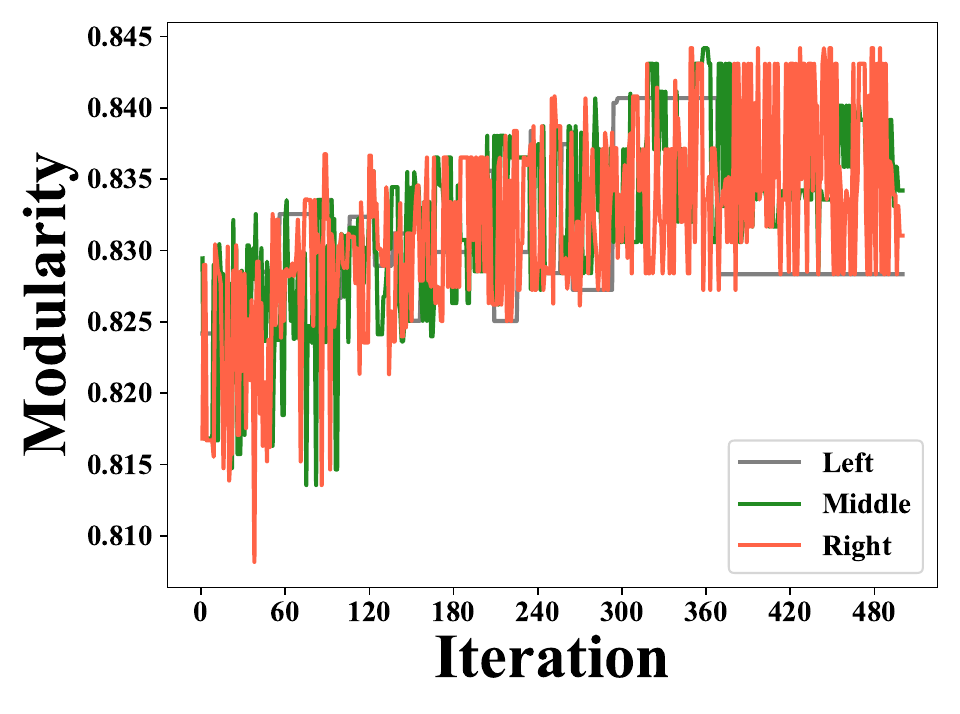}
			\end{minipage}%
		}%
		\\
		\subfigure[\textbf{Karate + LPA}]{
			\begin{minipage}[]{0.33\linewidth}
				\includegraphics[scale=0.33]{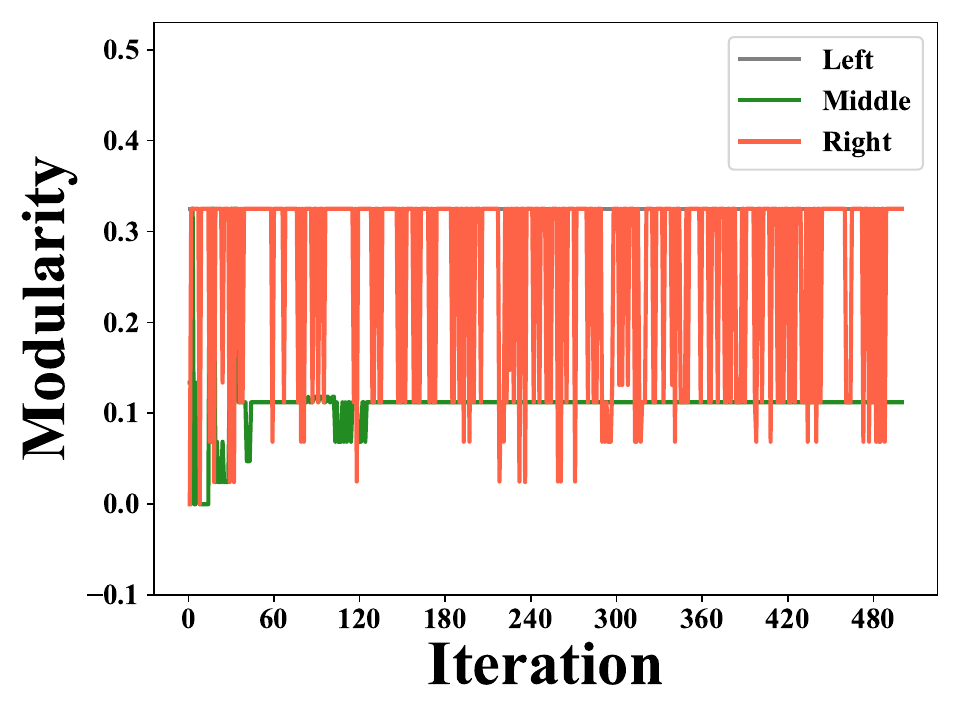}
			\end{minipage}%
		}%
		\subfigure[\textbf{Dolphins + LPA}]{
			\begin{minipage}[]{0.33\linewidth}
				\includegraphics[scale=0.33]{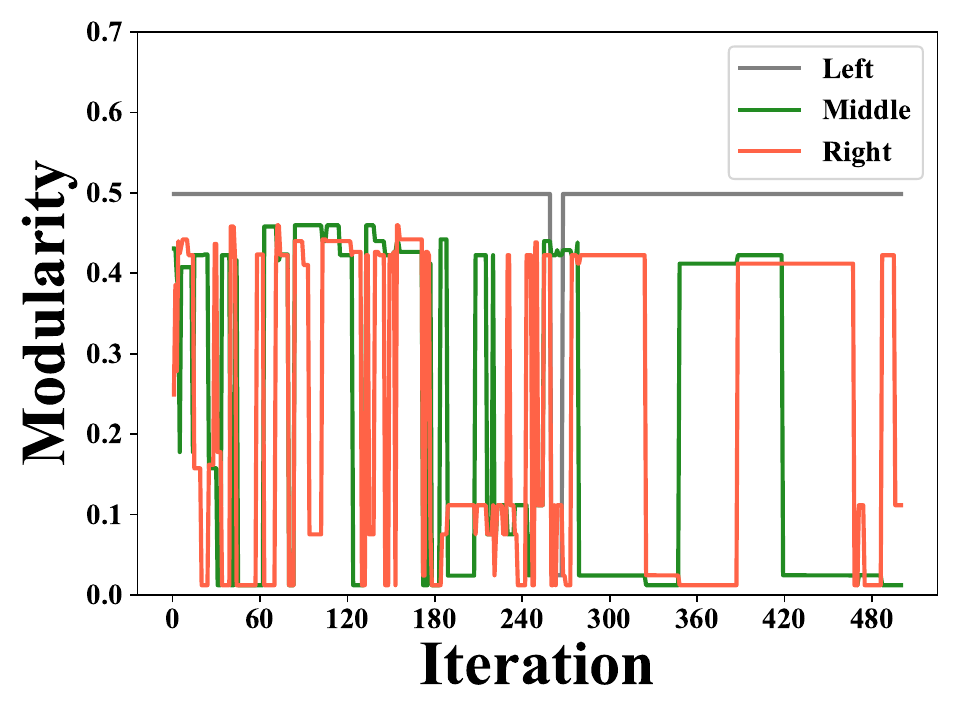}
			\end{minipage}%
		}%
		\subfigure[\textbf{Netscience + LPA}]{
			\begin{minipage}[]{0.33\linewidth}
				\includegraphics[scale=0.33]{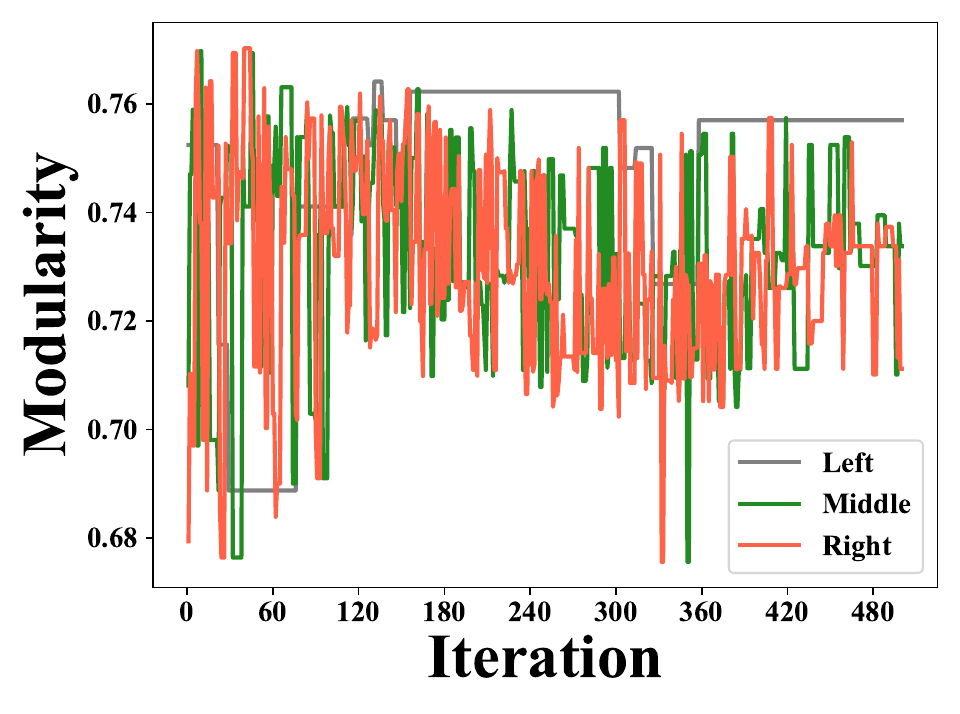}
			\end{minipage}%
		}%
		\centering
		\caption{Comparison of the change of modularity of different solutions of Pareto fronts obtained from UCD (MAX) with regard to the number of iterations under three detection algorithms on three datasets, where left, middle, and right indicate the farthest left, exact center, and farthest right solutions during each iteration.}
		\label{fig:mod}
	\end{figure*}

	\subsubsection{Relationship with Decrease of Modularity}
	Finally, we investigate the relationship between the traditional metric, the decrease of modularity, and the running iteration during the optimization of our UCDs. Similar to before, we select the farthest left solution, exact center solution, and farthest right solution as three representative solutions during each iteration, and plot their corresponding modularity values, as shown in Fig. \ref{fig:mod}. We can easily find that, with the increase of iteration (i.e., an increase of hyper volume under DARI and DAT), the value of modularity demonstrates significant fluctuations. The above finding also verifies our claims that the relationship between the decrease of detection effect and the decrease of modularity is not always linear. Sometimes a decrease of detection performance may lead to an increase of modularity. Thus, it may not be comprehensive enough to directly employ the decrease of modularity as the metric to evaluate the effect of community deception.

	\section{Conclusion}\label{sec:con}
    In contrast to community detection, which aims to cluster the densely connected nodes into the same group, community deception focuses on decreasing such clustering performance by injecting some unnoticeable perturbations. Specifically, the limitations of current deception metrics designed based on the classic modularity is first discussed. Then, we propose a new deception strategy, UCD, to solve the deception problem from a multi-objective optimization perspective, where both the deception performance and attack budget are considered simultaneously. Moreover, to further enhance the power of the proposed UCD, we develop two variant methods of the original UCD, i.e., UCD (MIN) and UCD (MAX), by considering the degree-biased and community-biased candidate node selection mechanisms in the mutation process. Comprehensive experiments demonstrate the superiority of the proposed UCDs, especially for their potential to enhance unnoticeability in community deception tasks.


	\bibliographystyle{IEEEtran}
	\bibliography{IEEEabrv, mybib}

\end{document}